\documentclass[a4paper,12pt]{article}
\pdfoutput=1

\usepackage{test,graphicx,subcaption,caption,framed}

\usepackage{xcolor}
\definecolor{pantone}{RGB}{1, 33, 105}

\usepackage{setspace} % load before hyperref so footnote links work
\usepackage{hyperref}
\hypersetup{colorlinks,linkcolor=blue,citecolor=blue,urlcolor=pantone}
\usepackage{doi}

\usepackage[style=ext-authoryear-comp,
sorting=nyt, % cite by name, year, title
sortcites=false, % prevent sorting citations
dashed=false, % don't use dash for repeated author
maxcitenames=2, % maximum names to cite in text
maxbibnames=99, % maximum names to list in reference
uniquelist=false,
uniquename=false,
giveninits=true, % first name initials
natbib, % use natbib commands
date=year % year only
]{biblatex}

\AtBeginRefsection{\GenRefcontextData{sorting=ynt}}
\AtEveryCite{\localrefcontext[sorting=ynt]}

\DeclareFieldFormat{pages}{#1} % suppress pp.
\renewbibmacro{in:}{\ifentrytype{article}{}{\printtext{\bibstring{in}\intitlepunct}}} % delete In:

\DeclareFieldFormat[article,inbook,incollection,inproceedings,patent,thesis,unpublished]{titlecase:title}{\MakeSentenceCase*{#1}} % change title to sentence case

\newbibmacro*{atoda:titlelink}[1]{%
  \ifhyperref
    {\iffieldundef{doi}
       {\iffieldundef{eprint}
          {\iffieldundef{url}
             {#1}
             {\href{\thefield{url}}{#1}}}
          {\href{https://arxiv.org/abs/\thefield{eprint}}{#1}}}
       {\href{https://doi.org/\thefield{doi}}{#1}}}
    {#1}}

\DeclareFieldFormat{title}{\usebibmacro{atoda:titlelink}{\mkbibemph{#1}}}
\DeclareFieldFormat
  [article,inbook,incollection,inproceedings,patent,thesis,unpublished]
  {title}{\usebibmacro{atoda:titlelink}{\mkbibquote{#1\isdot}}}

\renewbibmacro*{doi+eprint+url}{}

\usepackage[inline,shortlabels]{enumitem}
\setlist[enumerate,1]{label=(\roman*)}

\usepackage[margin = 1.25in]{geometry}
\usepackage{booktabs} %for nice tables

\newcommand{\spn}{\operatorname{span}}

\newcommand{\lr}[1]{\left[#1 \right]} % left right bracket
\newcommand{\lro}[1]{\left(#1 \right)} % left right parenthesis
\renewcommand{\hat}{\widehat}
\newcommand{\cH}{\mathcal{H}}
\newcommand{\cW}{\mathcal{W}}
\newcommand{\cX}{\mathcal{X}}
\newcommand{\cZ}{\mathcal{Z}}
\newcommand{\Expect}{\mathbf{E}} %expectation operator

\newcommand{\RI}{\texttt{RI}} % robust immunization
\newcommand{\HD}{\texttt{HD}\xspace} % high-order duration
\newcommand{\KRD}{\texttt{KRD}\xspace} % key rate duration
\newcommand{\LSHD}{\texttt{LSHD}\xspace}

\title{Robust Asset-Liability Management\thanks{We thank Bruno Biais, as well as seminar participants at HEC Paris, Norwegian School of Economics, Queen's University, UCSD, University of Hamburg, and the New York Camp Econometrics conference for valuable comments and feedback.}
}
\author{Tjeerd De Vries\thanks{Department of Finance, HEC Paris. Email: \href{mailto:de-vries@hec.fr}{de-vries@hec.fr}.} \and Alexis Akira Toda\thanks{Department of Economics, Emory University and Research Institute for Economics and Business Administration, Kobe University. Email: \href{mailto:alexis.akira.toda@emory.edu}{alexis.akira.toda@emory.edu}.}}

\numberwithin{equation}{section}
\numberwithin{prop}{section}

\begin{document}
\maketitle

\begin{abstract}

Financial institutions often cannot replicate long-dated liabilities with available bonds, especially when leverage and collateral constraints bind. We characterize the feasible portfolio that minimizes, to first order, the worst equity loss over a prespecified set of yield curve changes. The solution accommodates general deterministic liabilities and portfolio constraints and includes duration and convexity matching as special cases. The minimized loss measures the economic capital buffer needed to absorb the specified interest rate stress. Under an $\ell^2$ stress set, the portfolio is a constrained generalized least squares projection. A portfolio representation shows how additional stress directions can reduce reliance on unstable exact hedges. In U.S. Treasury applications, robust immunization delivers the best or nearly best funding performance for most liabilities and constraints. Simulated and historical dynamic exercises also show modest leverage, turnover, and transaction costs.

\medskip


\medskip

\end{abstract}

\section{Introduction}

Financial institutions invest assets to meet promised payments. Banks fund loans and securities with deposits and other debt. Insurers invest premiums to pay future claims. Defined benefit pension plans invest contributions to pay benefits. When interest rates change, they change the value of both assets and liabilities. The institution therefore cares about its equity or funding surplus, not about either side of the balance sheet in isolation. A loss of surplus can reduce its capacity to bear risk and can force it to raise capital, reduce lending, or sell assets \citep{froot1998risk,froot2008pricing,verani2024s}. The failures of Silicon Valley Bank\footnote{\href{https://www.ft.com/content/f9a3adce-1559-4f66-b172-cd45a9fa09d6}{Financial Times.}} and First Republic Bank\footnote{\href{https://www.economist.com/finance-and-economics/2023/05/03/what-the-first-republic-deal-means-for-americas-banks}{The Economist.}} in 2023 illustrate the consequences of large unhedged exposure to interest rates. The stress in the U.K. gilt market in 2022 shows that leverage and collateral requirements can constrain a hedge when rates move sharply. Evidence from Dutch pension funds also shows that margin calls on interest rate hedges can force sales of government bonds \citep{Breeden2022,Pinter2023,JansenKlinglerRanaldoDuijm2026}.

An institution could eliminate this risk if it could trade a zero coupon bond for every date on which it must make a payment. It could then replicate its liability cash flows exactly. Actual markets do not offer such a complete set of bonds. Traded maturities are sparse, and pension and insurance liabilities can extend beyond the longest available bond. Institutions also face limits on short positions and leverage, must post collateral, and incur trading costs. These features often prevent exact replication. They also make asset duration an incomplete measure of risk. The institution instead needs to know how well the best feasible portfolio can protect the value of its assets relative to its liabilities.

Classical immunization provides the natural starting point. \citet{FisherWeil1971} show how value and duration matching protect a portfolio against parallel shifts in the yield curve. Later papers extend immunization to multiple liability payments and broader families of yield curve changes, and study maxmin or minimax portfolios when exact immunization fails.\footnote{See, \eg, \citet{BierwagKaufmanToevs1983}, \citet{shiu1988immunization}, \citet{balbas1998can}, \citet{BarberCopper1998}, \citet{Balbas2002}, and \citet{BalbasRomera2007}.} Higher order duration and factor methods allow more varied changes in the yield curve, but exact matching can require large offsetting long and short positions \citep{Mantilla-Garcia2022}. We build on this literature by showing that, to first order, maximizing equity in the worst case is equivalent to minimizing maximum sensitivity. This connection yields a tight local loss bound under general closed portfolio constraints. For common stress sets and linear constraints, it also yields tractable solutions. We call the method \emph{robust immunization} because we do not assume a particular change in the yield curve. Within a stress set (the class of yield curve changes against which we seek protection), we allow the curve to move in any direction, including changes in its level, slope, and curvature. Given the liability, the current yield curve, the available bonds, and the portfolio constraints, we choose the feasible portfolio that maximizes equity under the most adverse change in that stress set, to first order.

We first characterize the protection that any feasible portfolio can provide. Under weak conditions, we prove that a robust immunizing portfolio exists and minimizes the worst equity loss to first order (Proposition \ref{prop:minmax} and Theorem \ref{thm:RobIm}). To first order, the resulting loss equals the product of three terms: the present value of the liability, the size of the yield curve stress, and a measure of the exposure that remains after the institution chooses its best feasible hedge. We interpret this loss as the minimum capital buffer needed to absorb the specified stress, where ``capital buffer'' has an economic rather than regulatory meaning. The remaining exposure depends on the available bonds, the liability cash flows, the possible changes in the yield curve, and the portfolio constraints. Adding possible changes cannot reduce the required buffer, while expanding the set of feasible portfolios cannot increase it. Thus, we measure interest rate risk at the level of the entire balance sheet rather than the asset portfolio alone.

We next show that robust immunization includes familiar hedging methods as special cases. If the yield curve can move only in parallel, our solution reduces to value and duration matching. Polynomial perturbations produce duration, convexity, and higher order duration matching. Factor loadings produce factor duration matching. When the liability pays on one date and the portfolio has no short positions, our global funding result recovers the theorem of \citet{FisherWeil1971} (Section \ref{subsec:main_lit}). We can also require the portfolio to match important directions, such as level or duration risk, exactly and hedge the remaining changes robustly (Theorem \ref{thm:RobIm1}). For implementation, we use transformed Chebyshev polynomials. Their first three loadings are constant, linear, and quadratic in maturity, so they mimic the level, slope, and curvature movements that account for most changes in the yield curve.

We also derive a simple way to compute the portfolio. Under the baseline $\ell^\infty$ stress set, we solve a finite convex optimization problem. Under an $\ell^2$ stress set and linear constraints, we obtain a constrained generalized least squares solution in closed form (Proposition \ref{prop:RobIml2}). We then express this solution as a combination of portfolios that match different sets of shocks exactly. The weights in this representation need not be positive. We provide a sufficient condition under which they are positive and show that nearly singular matching systems then receive little weight. This result gives a precise sense in which additional shock directions can reduce the influence of unstable exact hedges. It also helps explain why the robust portfolios use moderate leverage in our applications.

We evaluate the method with U.S. Treasury yield curves and five liability structures, including a pension liability. We use data from 1985--1995 to choose the number of basis functions and reserve 1996--2023 for evaluation. We compare robust immunization with least squares higher order duration matching and key rate duration matching. We impose the same three sets of constraints on each method: no short positions, a gross leverage cap of two, and a collateral budget based on supervisory haircuts. Our static exercise uses realized changes in the yield curve over horizons from one to one hundred trading days. Our dynamic exercise rebalances the portfolio quarterly for ten years and includes proportional transaction costs. We conduct the dynamic exercise on 5,000 simulated paths and on 72 overlapping windows from the historical yield curves.

In the static exercise, we find that robust immunization performs consistently across different liability structures when the required matching constraints are feasible. Under the leverage cap, robust duration and convexity immunization has the best or nearly best downside performance for each of the four liabilities for which it is feasible. For the pension liability, its 95th percentile of thirty day underfunding is 0.15\%, compared with 0.34\% for least squares higher order duration matching and 2.27\% for key rate duration matching. The exception is a liability that pays only after year twenty. Its duration exceeds the maturity of the longest available bond, so the leverage cap prevents duration and convexity matching. This exception shows that portfolio constraints can limit the protection that any method can provide. More generally, the constraints often improve realized performance by removing large offsetting positions, but they can also make an otherwise desirable match infeasible.

In the dynamic exercises, we find that small, persistent mismatches matter. On the simulated paths, robust immunization produces the lowest mean and mean squared terminal errors for both liabilities that we study. It also uses less leverage and turnover than the closest feasible benchmark. On the historical paths, its advantage grows because yields moved persistently over long periods. For the pension liability, robust immunization has a mean terminal error of 0.64\% and a 99th percentile of 1.78\% across the historical ten year windows. In contrast, methods with a persistent signed duration gap accumulate errors in the same direction as yields trend. A portfolio with little turnover can therefore remain risky if it repeatedly carries the same mismatch. Transaction costs do not change the rankings because robust immunization trades less than the closest benchmark.

Our theory studies small changes in yields and fixed liability cash flows. Our empirical exercises therefore assess practical performance rather than global optimality. We do not model stochastic liabilities, credit risk, market liquidity, or the endogenous choice of trading costs. Within this scope, we show that institutions should measure interest rate risk by the loss of the best feasible asset and liability position under the worst specified change in yields, rather than by the sensitivity of the assets alone.

\subsection{Related literature}\label{subsec:litreview}

Our paper first contributes to the literature on financial intermediary risk management. \citet{froot1998risk} study how internal capital and financing frictions shape risk management and capital budgeting. \citet{froot2008pricing} analyze how constrained intermediaries price the risks that they bear. \citet{verani2024s} study how intermediary balance sheets affect annuity markets. We take prices, liabilities, and constraints as given instead of modeling their equilibrium determination. We characterize the portfolio that minimizes the local loss of net worth under the worst specified yield change and show how that loss depends on the liabilities, available bonds, possible yield changes, and portfolio constraints.

A second literature studies bond immunization. \citet{Macaulay1938}, \citet[pp.~184--188]{Hicks1939}, and \citet{Samuelson1945} identify duration as the sensitivity of a bond to parallel changes in interest rates, and \citet{Redington1952} proposes matching the duration of assets and liabilities. \citet{FisherWeil1971} establish a global duration matching guarantee. \citet{BierwagKhang1979} give this strategy a maxmin interpretation, which \citet{prisman1986immunization} extends to a setting with taxes. \citet{BierwagKaufmanToevs1983} and \citet{shiu1988immunization} study multiple liability payments. \citet{FongVasicek1984}, \citet{bowden1997generalising}, and \citet{balbas1998can} study richer changes in interest or forward rates. \citet{BarberCopper1998} minimize maximum portfolio sensitivity. \citet{Balbas2002} compute maxmin portfolios when exact immunization is infeasible, and \citet{BalbasRomera2007} formulate constrained maximin hedges for multiple liabilities in Banach spaces. We connect these strands by deriving a minimax sensitivity problem from the local maxmin equity problem. For general closed constraint sets, we prove existence and derive a tight error bound. Under common stress sets and linear constraints, we obtain tractable solutions and portfolio representations. Section \ref{subsec:main_lit} makes the comparison precise.

Several papers develop hedges based on multiple factors or higher order duration. \citet{ChambersCarletonMcEnally1988}, \citet{NawalkhaLacey1988}, and \citet{PrismanShores1988} use polynomial approximations and higher order duration measures. \citet{Ho1992} proposes key rate duration. \citet{LittermanScheinkman1991} identify level, slope, and curvature factors, and \citet{Willner1996} applies the corresponding factor durations to the management of assets and liabilities. Using simulations and U.S. Treasury data, \citet{Agca2005} and \citet{CarcanoDallo2011} show that portfolio construction, holding periods, transaction costs, and model error can matter more than the chosen risk measure. \citet{Mantilla-Garcia2022} show that regularization can improve interest rate hedges with unstable weights. Our generalized least squares and elemental portfolio representations identify conditions under which an expanded set of shock directions reduces the influence of unstable exact matching portfolios.

Finally, our paper relates to robust asset and liability management and to the construction of term structure factors. \citet{GulpinarPachamanova2013} use robust optimization to study multiperiod pension asset and liability management when investment opportunities vary over time. We instead choose a bond portfolio against a stress set of yield curve changes and derive local analytical guarantees under portfolio constraints. Our maxmin criterion is also related to the maxmin expected utility criterion of \citet{GilboaSchmeidler1989}. \citet{filipovic2025stripping,filipovic2026shrinking} construct discount curves and term structure factors without conventional covariance estimation. They seek to estimate the yield curve and explain the risk premia of Treasury returns. We seek to hedge a specified liability subject to portfolio constraints. We use their basis functions and discount curve database to check the robustness of our results.

The remainder of the paper proceeds as follows. Section~\ref{sec:prob} formulates the asset and liability problem. Section~\ref{sec:main} derives the robust immunization theorem and relates it to classical immunization. Section~\ref{sec:extensions} develops the $\ell^2$ representation, the regularization interpretation, and the implementation. Section~\ref{sec:eval} presents the static and dynamic evaluations. Section~\ref{sec:conclusion} concludes.

\section{Problem statement}\label{sec:prob}

\subsection{Classical immunization}

We start the discussion with a brief review of classical immunization. Consider cash flows (liabilities) $f_1,\dots,f_N>0$ paid out at time $t_1<\dots<t_N$. Assuming a constant (continuously-compounded) interest rate $r$, the present value of liabilities is
\begin{equation}
    P\coloneqq \sum_{n=1}^N \e^{-rt_n}f_n. \label{eq:presentvalue}
\end{equation}
Because each term in the sum \eqref{eq:presentvalue} is decreasing in $r$, so is the present value $P$. Therefore we may define the interest-rate sensitivity of liability by
\begin{equation}
    D\coloneqq -\frac{\partial \log P}{\partial r}=-\frac{1}{P}\frac{\partial P}{\partial r}=\frac{1}{P}\sum_{n=1}^N t_n\e^{-rt_n}f_n>0. \label{eq:duration}
\end{equation}
Intuitively, $D$ is the percentage change in present value with respect to a one percentage point change in the interest rate. The quantity $D$ in \eqref{eq:duration} is called the \emph{duration} of the cash flow $(f_1,\dots,f_N)$ because it can be interpreted as the average time to payment. To see why, define the weight $w_n=\e^{-rt_n}f_n/P$, which by \eqref{eq:presentvalue} is the share of present value attributable to the payment at $t_n$. Since $w_n>0$ and $\sum_{n=1}^N w_n=1$, the definition of duration \eqref{eq:duration} implies $D=\sum_{n=1}^N w_nt_n$ is indeed a weighted average of time to payment.

In general, the interest rate need not be constant across different maturities. If $y(t)$ denotes the pure yield for maturity $t$ (so by definition the price of a zero-coupon bond with face value 1 and maturity $t$ is $\e^{-y(t)t}$), then the present value and duration of the cash flows become
\begin{align*}
    P(y)&=\sum_{n=1}^N \e^{-y(t_n)t_n}f_n,\\
    D(y)&=- \frac{1}{P(y)}\lim_{\Delta\to 0}\frac{P(y+\Delta)-P(y)}{\Delta}=\frac{1}{P(y)}\sum_{n=1}^N t_n\e^{-y(t_n)t_n}f_n,
\end{align*}
respectively. Here the duration is the sensitivity of the present value with respect to an infinitesimal parallel shift in the yield curve. The idea of classical \emph{immunization} is to match the duration of asset and liability so that equity (asset minus liability) is insensitive to yield curve shifts \citep{Macaulay1938,Samuelson1945,Redington1952,FisherWeil1971,BierwagKhang1979}. Below, we generalize classical immunization to overcome the limitations discussed in the introduction. 

\subsection{Robust immunization problem}

We now turn to the description of the robust immunization problem. Time is continuous and denoted by $t\in [0,T]$, where $T>0$ is the planning horizon. There are finitely many bonds available for trade indexed by $j=1,\dots,J$, where $J\ge 2$. The cumulative payout of bond $j$ is denoted by the (weakly) increasing function $F_j:[0,T]\to \R_+$. For instance, if bond $j$ is a zero-coupon bond with face value normalized to 1 and maturity $t_j$, then
\begin{equation}
    F_j(t)=\begin{cases*}
        0 & if $0\le t<t_j$, \\
        1 & if $t_j\le t\le T$.
    \end{cases*}\label{eq:zero_coupon}
\end{equation}
Similarly, if bond $j$ continuously pays out coupons at rate $c_j>0$ and has zero face value, then $F_j(t)=c_jt$ for $0\le t\le T$.

The fund manager seeks to immunize future cash flows against interest rate risk by forming a portfolio of bonds $j=1,\dots,J$. Let $F:[0,T]\to \R_+$ be the cumulative cash flow to be immunized and $y:[0,T]\to \R$ be the yield curve, which the fund manager takes as given. The present value of cash flows is given by the Riemann-Stieltjes integral
\begin{equation}
    \int_0^T \e^{-ty(t)}\diff F(t).\label{eq:PV}
\end{equation}
Because the expression $ty(t)$ appears frequently, it is convenient to introduce the notation $x(t)\coloneqq ty(t)$. Note that by the definition of the instantaneous forward rate $f(t)$ at term $t$, we have
\begin{equation}
    x(t)=\int_0^t f(u)\diff u. \label{eq:xt}
\end{equation}
Because $x$ is the integral of forward rates, we refer to it as the \emph{cumulative discount rate}. Using $x$, we can rewrite the present value of cash flows \eqref{eq:PV} as
\begin{equation}
    P(x)\coloneqq \int_0^T \e^{-x(t)}\diff F(t),\label{eq:Px}
\end{equation}
which is a functional of $x$. We can define the price $P_j(x)$ of bond $j$ analogously. The fund manager's problem is to approximate $P(x)$ using a linear combination of bonds $\set{P_j(x)}_{j=1}^J$ so that the approximation is robust against perturbations to the yield curve $y$ (and hence the cumulative discount rate $x$).

To define this portfolio choice problem, let $\cZ\subset \R^J$ be the set of admissible portfolios and let $\cH$ be the \emph{stress set}, which contains the perturbations to the cumulative discount rate against which the fund manager seeks protection. We specify both sets later. For portfolio $z\in \cZ$ and perturbation $h\in \cH$, define the ``asset'' by
\begin{equation*}
    V(z,x+h)\coloneqq \sum_{j=1}^J z_j P_j(x+h)
\end{equation*}
and the ``equity'' by asset minus liability
\begin{equation}
    E(z,x+h)\coloneqq V(z,x+h)-P(x+h)=\sum_{j=1}^J z_j P_j(x+h)-P(x+h). \label{eq:equity}
\end{equation}
We assume the fund manager has a maxmin preference and seeks to solve
\begin{equation}
    \sup_{z\in \cZ}\inf_{h\in \cH} E(z,x+h).\label{eq:maxmin}
\end{equation}
The interpretation of the maxmin problem \eqref{eq:maxmin} is as follows. Given the portfolio $z\in \cZ$, nature chooses the most adversarial perturbation $h\in \cH$ to minimize equity. The fund manager chooses the portfolio $z$ that guarantees the highest equity under the worst possible perturbation. We refer to this problem as the \emph{robust immunization problem}. Maxmin preferences are common in the literature on bond portfolio choice; see, \eg, the structural models of \citet{gagliardini2008ambiguity} and \citet[Appendix B]{vayanos2021preferred}. These preferences can also be interpreted as those of a policymaker who seeks to prevent bankruptcy due to the societal costs associated with the collapse of a large financial institution. 

The objective function in \eqref{eq:maxmin} can be microfounded using the ambiguity-averse preferences of \citet{GilboaSchmeidler1989}, assuming a set of degenerate priors, one for each admissible shock.\footnote{It can also be interpreted as a Wald maxmin criterion.} This prior set may be considered overly dogmatic and, from a technical standpoint, is not convex. In Appendix \ref{app:convex}, we show that the maxmin problem in Theorem \ref{thm:RobIm} is unchanged if the set of priors is extended to all probability measures supported on the stress set. This is a convex set of priors, which aligns precisely with the framework of \citet{GilboaSchmeidler1989}.

\subsection{Assumptions}

The maxmin problem \eqref{eq:maxmin} is intractable because we have not yet specified the portfolio set $\cZ$ or the stress set $\cH$, and the objective function is nonlinear (not even convex) in $h$. We thus impose several assumptions to make progress.

\begin{asmp}[Discrete payouts]\label{asmp:payouts}
The bonds and liability pay out on finitely many dates, whose union is denoted by $\set{t_n}_{n=1}^N\subset (0,T]$.
\end{asmp}

Assumption \ref{asmp:payouts} always holds in practice. Under this assumption, each $F_j$ is a step function with discontinuities at points contained in $\set{t_n}_{n=1}^N$, and integrals of the form \eqref{eq:Px} reduce to summations.

\begin{asmp}[Portfolio constraint]\label{asmp:Z}
The set of admissible portfolios $\cZ\subset \R^J$ is nonempty and closed. Furthermore, all $z\in \cZ$ satisfy value matching:
\begin{equation}
    P(x)=\sum_{j=1}^J z_jP_j(x). \label{eq:value_match}
\end{equation}
\end{asmp}

Value matching \eqref{eq:value_match} is merely a normalization to make the initial equity (asset minus liability) equal to 0. This assumption is standard in the immunization literature (see, for example, \citet{BierwagKhang1979}). Note that Assumption \ref{asmp:Z} allows for short sale constraints either on the entire portfolio or on individual bonds. The latter constraint may arise if the portfolio manager is concerned about short selling illiquid long-term bonds. 

We now specify the space of cumulative discount rates and their perturbations. Let the yields be in $C[0,T]$, the vector space of continuous functions on $[0,T]$ endowed with the supremum norm denoted by $\norm{\cdot}_\infty$.\footnote{As we use several different norms in this paper, we use subscripts to distinguish them. An example is the $\ell^p$ norm on $\R^J$ for $p=1,2$, which we denote by $\norm{\cdot}_p$.} By the definition of the cumulative discount rate, we obtain that $x:[0,T]\to \R$ defined by $x(t) = t y(t)$ is continuous with $x(0)=0$. We thus define the space of cumulative discount rates by
\begin{equation}
    \cX=\set{x\in C[0,T]:x(0)=0}.\label{eq:cX}
\end{equation}
Lemma \ref{lem:cX_Banach} shows that $\cX$ is a Banach space. The next assumption allows us to approximate any element $x\in \cX$, which is important for reducing the problem to a finite dimension.

\begin{asmp}\label{asmp:H}
There exists a sequence $\set{h_i}_{i=1}^\infty\subset \cX$ whose linear span is dense in $\cX$ and such that, for each $I\le N$, the $I\times N$ matrices $H\coloneqq (h_i(t_n))$ and $G\coloneqq (h_i(t_n)/t_n)$ have full row rank.
\end{asmp}

We refer to each $h_i$ as a \emph{basis function}. The matrices $H$ and $G$ will be used to approximate the discount rate and yield curve at the payout dates.\footnote{Recall that the yield curve satisfies $y(t) = x(t)/t$.} Assumption \ref{asmp:H} says that the basis functions are linearly independent when evaluated on the payout dates. We impose this assumption to avoid portfolio indeterminacy. In practice, we can always ensure that $H$ and $G$ have full row rank by removing redundant basis functions if necessary. A typical example satisfying Assumption \ref{asmp:H} is to let $h_i$ be a polynomial of degree $i$ with $h_i(0)=0$ (Lemma \ref{lem:poly}). We now consider several examples of perturbations to the yield curve (or cumulative discount rate) that can arise.

\begin{exmp}[Classical Immunization]
The setting in classical immunization corresponds to $I=1$ and $h_1(t)=t$ (hence $h_1(t)/t=1$), which implies that the perturbations to the yield curve are restricted to parallel shifts.
\end{exmp}

\begin{exmp}[Vasicek Model]\label{exmp:vasicek}
In the \citet{vasicek1977equilibrium} model, the change in yields from time $s$ to $s'$ is given by 
\begin{equation}\label{eq:vasicek}
y_{s'}(t) - y_s(t) = (r_{s'} - r_s) \frac{1- \e^{-a t}}{a t}  \eqqcolon h(t)/t,
\end{equation}
where $r_s$ is the spot rate that solves the stochastic differential equation
\begin{equation*}
\diff r_s = a(b- r_s) \diff s + \sigma  \diff W_s.
\end{equation*}
A similar, though more complicated expression for the yield changes holds in the equilibrium model of \citet[Appendix B]{vayanos2021preferred}. Equation \eqref{eq:vasicek} implies that the model does not allow for parallel shifts of the yield curve. As we shall see, classical immunization is therefore never a maxmin strategy.

\begin{exmp}[Principal Components]\label{exmp:pca}
Similar to \citet{diebold2006forecasting}, we can specify a 3-factor model for the yield curve at time $s$ by:
\begin{equation*}
y_s(t) = \beta_{1s} h_1(t)/t + \beta_{2s} h_2(t)/t + \beta_{3s} h_3(t)/t,
\end{equation*}
where $h_i(t)/t$ for $i=1,2,3$ represent the loading on the level, slope, and curvature factors (see Figure \ref{fig:pca}). For example, an increase in $\beta_{2s}$ will increase short yields more relative to long yields, thereby changing the slope of the yield curve. Similarly, an increase in $\beta_{3s}$ primarily increases medium yields around the two-year maturity, while short and long yields remain unaffected. This leads to an increase in the curvature of the yield curve. In this setting, perturbations to the yield curve are restricted to changes in level, slope, and curvature. \citet{LittermanScheinkman1991} find that these components are the main drivers of changes in the yield curve, although recent work of \citet{CrumpGospodinov2022} suggests that the factor dimension may be larger. 

\begin{figure}[!htb]
    \centering
    \includegraphics[width=0.7\linewidth]{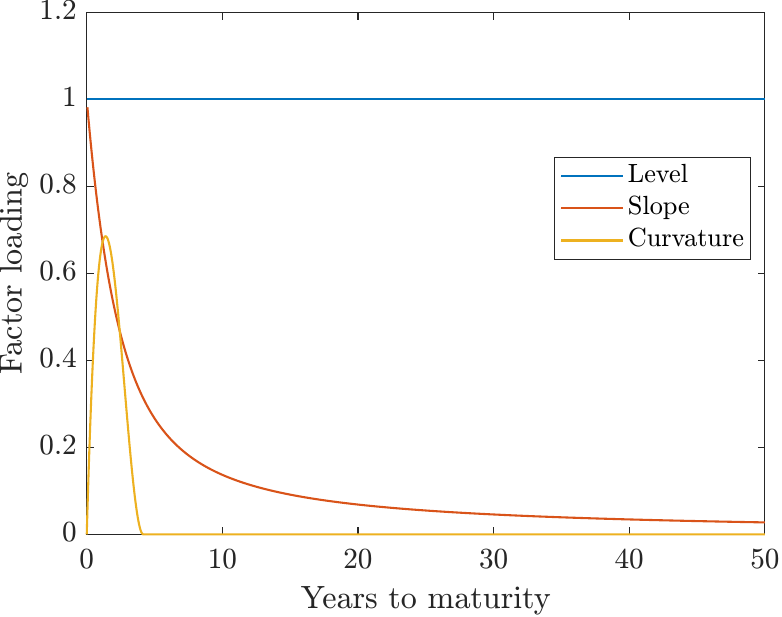}
    \caption{Factor loadings on the level, slope, and curvature factors in the \citet{diebold2006forecasting} model.}
    \label{fig:pca}
\end{figure}
    
\end{exmp}

Finally, we specify the baseline stress set for perturbations to the cumulative discount rate. For any $\Delta>0$, define
\begin{equation}
    \cH_I(\Delta)\coloneqq \set*{h\in \spn\set{h_i}_{i=1}^I: (\forall n) \abs{h(t_n)/t_n} \le \Delta}.\label{eq:cH_Delta}
\end{equation}
Because $h$ is a perturbation to the cumulative discount rate, choosing $h\in \cH_I(\Delta)$ amounts to allowing the yields to change by at most $\pm \Delta$ within the span of the first $I$ basis functions. With this specification of the stress set, the robust immunization problem \eqref{eq:maxmin} becomes
\begin{equation}
    \sup_{z\in \cZ}\inf_{h\in \cH_I(\Delta)} E(z,x+h).\label{eq:maxmin_I}
\end{equation}
\end{exmp}

\section{Solving robust immunization problem}\label{sec:main}

In this section we solve the maxmin problem \eqref{eq:maxmin_I} in the limit as $\Delta\downarrow 0$. The resulting optimal portfolio is therefore expected to perform well when shocks to the yield curve are small. This is the price we pay for deriving a portfolio that is maxmin optimal against a broad class of yield curve perturbations and portfolio constraints, rather than restricting attention to specific shifts and ruling out short sales, both of which are strong assumptions. Whether yield curve shocks are sufficiently small in practice for the maxmin solution to perform well is an empirical question, which we analyze in Section~\ref{sec:eval}.\footnote{Analogous situations often arise in econometrics, for example in the weak-instrument literature. 
In that setting, the finite-sample behavior of the IV estimator is often better approximated by an asymptotic regime in which the correlation between the regressor and the instrument drifts to zero, rather than by standard asymptotics that assume a fixed correlation bounded away from zero.}

\subsection{Robust immunization}\label{subsec:main_RobIm}

As the set of cumulative discount rates $\cX$ forms an infinite-dimensional vector space, we employ tools from functional analysis to analyze how prices change in response to perturbations in the discount rate $h \in \cH_I(\Delta)$. We assess the price change induced by an arbitrary shift in the cumulative discount rate by using the Gateaux differential of $P(x)$:
\begin{equation}
    \delta P(x;h)\coloneqq \lim_{\alpha\to 0}\frac{1}{\alpha}(P(x+\alpha h)-P(x))=-\int_0^T \e^{-x(t)}h(t)\diff F(t).\label{eq:gateaux}
\end{equation}

\begin{rem}
The operator $h\mapsto \delta P(x;h)$ defined by \eqref{eq:gateaux} is a bounded linear operator from $\cX$ to $\R$ (Lemma \ref{lem:bounded_linear_operator}), which is called the Fr\'echet derivative and denoted by $P'(x)$. Thus by definition $P'(x)h=\delta P(x;h)$. In broad terms, $P'(x)h$ quantifies the first-order effect on price when the cumulative discount rate curve is perturbed by $h$.
\end{rem}

We construct a solution to the maxmin problem \eqref{eq:maxmin_I} by assessing the sensitivities of the asset and liability to perturbations in specific directions $h$. Specifically, given the basis functions $\set{h_i}_{i=1}^I$ and bonds $j=1,\dots,J$, we define the \emph{sensitivity vector} $b=(b_i)\in \R^I$ of liabilities by
\begin{equation}
    b_i\coloneqq -\frac{P'(x)h_i}{P(x)}=-\frac{\delta P(x;h_i)}{P(x)}=\frac{1}{P(x)}\int_0^T \e^{-x(t)}h_i(t)\diff F(t). \label{eq:bi}
\end{equation}
Note that the duration \eqref{eq:duration} corresponds to the special case of $h_i(t)=t$, and therefore $b_i$ is a generalization. Intuitively, each entry $b_i$ represents the sensitivity of liability to a perturbation evaluated at $h = h_i$. Similarly, we define the \emph{sensitivity matrix} $A = (a_{ij}) \in \R^{I \times J}$ by
\begin{equation}
    a_{ij}\coloneqq -\frac{P_j'(x)h_i}{P(x)}=-\frac{\delta P_j(x;h_i)}{P(x)}=\frac{1}{P(x)}\int_0^T \e^{-x(t)}h_i(t)\diff F_j(t).\label{eq:aij}
\end{equation}
Division by $P(x)$ is merely a normalization to make $a_{ij}$ unit-free. Again, each entry $a_{ij}$ represents the sensitivity of bond $j$ (with $F = F_j$) to a perturbation evaluated at $h = h_i$.

If $h\in \cH_I(\Delta)$ in \eqref{eq:cH_Delta}, so $h=\Delta \sum_{i=1}^I w_ih_i$ for some $w\in \R^I$ (the coefficient $\Delta$ is to make $w$ scale-free), then using the definition of equity \eqref{eq:equity} and noting that $E(z,x)=0$ by Assumption \ref{asmp:Z}, we obtain the sensitivity of equity
\begin{equation}
    \lim_{\Delta\to 0}\frac{1}{\Delta P(x)}E(z,x+h)=-\seq{w,Az-b}, \label{eq:E_gateaux}
\end{equation}
where $\seq{\cdot,\cdot}$ denotes the inner product. Hence, the change in equity following an infinitesimal perturbation in the discount rate is governed by the Fr\'echet derivatives of the asset and liability. For $h=\Delta\sum_{i=1}^I w_ih_i\in \cH_I(\Delta)$, the coefficients $(w_i)$ need to satisfy particular restrictions. Using the definition of the matrix $G$ in Assumption \ref{asmp:H} and \eqref{eq:cH_Delta}, it is straightforward to show $h=\Delta\sum_{i=1}^I w_ih_i\in \cH_I(\Delta)$ if and only if $G'w\in [-1,1]^N$. This observation, together with \eqref{eq:maxmin_I} and \eqref{eq:E_gateaux} motivate us to define the set
\begin{equation}
    \cW\coloneqq \set{w\in \R^I: G'w\in [-1,1]^N}\label{eq:cW}
\end{equation}
and the minmax problem
\begin{equation}
    V_I(\cZ) \coloneqq \inf_{z\in \cZ}\sup_{w\in \cW}\seq{w,Az-b}. \label{eq:VIZ}
\end{equation}

The next proposition establishes the existence of a solution to the minmax problem \eqref{eq:VIZ}. Before stating this result, it is convenient to introduce notation for the value matching constraint, which is always assumed to hold (Assumption \ref{asmp:Z}). Specifically, set $h_0\equiv 1$ and define $a_{0j}$ using \eqref{eq:aij}. Define the $1\times J$ vector $a_0\coloneqq (a_{0j})$ and the $(I+1)\times J$ matrix and $(I+1)\times 1$ vector
\begin{equation}
    A_+\coloneqq \begin{bmatrix}
        a_0 \\ A
    \end{bmatrix} \quad \text{and} \quad 
        b_+\coloneqq \begin{bmatrix}
        1 \\ b
    \end{bmatrix}.\label{eq:Abplus}
\end{equation}

In what follows, proofs are deferred to Appendix \ref{sec:proof}.

\begin{prop}[Minmax]\label{prop:minmax}
Suppose Assumptions \ref{asmp:payouts}--\ref{asmp:H} hold, $I\ge J-1$, and $A_+$ in \eqref{eq:Abplus} has full column rank. Then the following statements are true.
\begin{enumerate}
    \item There exists $(z^*,w^*)\in \cZ\times \cW$ that achieves the minmax value \eqref{eq:VIZ}.
    \item $V_I(\cZ)\ge 0$, and $z\in \cZ$ achieves $V_I(\cZ)=0$ if and only if $A_+z=b_+$.
\end{enumerate}
\end{prop}

The solution $z$ to the minmax problem \eqref{eq:VIZ} depends on the basis functions $\set{h_i}_{i=1}^I$ only through its span and it is immaterial how we parameterize these functions.

\begin{prop}[Basis invariance]\label{prop:invariance}
Let everything be as in Proposition \ref{prop:minmax} and $\cZ^*$ be the set of solutions $z^*\in \cZ$ to the minmax problem \eqref{eq:VIZ}. Then $V_I(\cZ)$ and $\cZ^*$ depend on the basis functions $\set{h_i}_{i=1}^I$ only through its span.
\end{prop}

Proposition \ref{prop:minmax} assumes that $A_+$ in \eqref{eq:Abplus} has full column rank, which holds under weak conditions. If the cumulative payouts of bonds $\set{F_j}$ and the basis functions $\set{h_i}$ are linearly independent, the matrix $A_+$ generically has full column rank and therefore a solution $(z,w)\in \cZ\times \cW$ to the minmax problem \eqref{eq:VIZ} generically exists. In Appendix \ref{app:full_column_A} we make this statement more precise.

Before presenting our first main result, we introduce one last piece of notation. For any bond portfolio $z\in \cZ$, define the portfolio share $\theta=(\theta_j)\in \R^J$ by
\begin{equation}
    \theta_j\coloneqq z_jP_j(x)/P(x). \label{eq:theta}
\end{equation}
Assuming  value matching (Assumption \ref{asmp:Z}), the portfolio share $\theta$ satisfies $\sum_{j=1}^J\theta_j=1$. Therefore the $\ell^1$ norm $\norm{\theta}_1=\sum_{j=1}^J\abs{\theta_j}$ satisfies $\norm{\theta}_1=1$ if and only if $\theta_j\ge 0$ for all $j$, and $\norm{\theta}_1>1$ is equivalent to $\theta_j<0$ for some $j$. Thus $\norm{\theta}_1$ can be interpreted as a measure of leverage, which we refer to as the \emph{gross leverage}. 

\begin{thm}[Robust immunization]\label{thm:RobIm}
Let everything be as in Proposition \ref{prop:minmax} and $\cH_I(\Delta)$ be as in \eqref{eq:cH_Delta}. Then the following statements are true.
\begin{enumerate}
\item The guaranteed equity satisfies
\begin{equation}
    \lim_{\Delta\downarrow 0}\frac{1}{\Delta}\sup_{z\in \cZ}\inf_{h\in \cH_I(\Delta)} E(z,x+h)=-P(x) V_I(\cZ).\label{eq:maxmin0}
\end{equation}
\item Letting $z^*\in \cZ$ be the solution to the minmax problem \eqref{eq:VIZ} and $\theta=(\theta_j)\in \R^J$ be the corresponding portfolio share defined by \eqref{eq:theta}, then
\begin{equation}
    \sup_{h\in \cH_I(\Delta)}\abs{E(z^*,x+h)}\le \Delta P(x)\left(V_I(\cZ)+\frac{1}{4}\Delta T^2\e^{\Delta T}(1+\norm{\theta}_1)\right).\label{eq:RobIm_error_bd}
\end{equation}
\end{enumerate}
\end{thm}

Theorem \ref{thm:RobIm} has several implications. First, \eqref{eq:maxmin0} shows that, to first order, the guaranteed equity is exactly $-\Delta P(x) V_I(\cZ)$ when yields are perturbed by at most $\pm \Delta$ within the span of the basis functions. The minmax value $V_I(\cZ)$ has a natural interpretation and is the answer to the following question: ``if yields change by at most one percentage point, what is the largest percentage point decline in the portfolio value?'' The maxmin formula \eqref{eq:maxmin0} provides an exact characterization of the worst-case outcome, and the number $V_I(\cZ)$ is the value of the minmax problem \eqref{eq:VIZ}.\footnote{In Appendix \ref{app:convex}, we show that the maxmin value is unchanged if the set of priors is extended to all probability measures supported on the stress set, which fits within the ambiguity-averse framework of \citet{GilboaSchmeidler1989}.} Second, the error estimate \eqref{eq:RobIm_error_bd} shows that the solution $z^*\in \cZ$ to the minmax problem \eqref{eq:VIZ} achieves the lower bound in \eqref{eq:maxmin0}, to first order. In this sense $z^*$ is an optimal portfolio, which we refer to as the \emph{robust immunizing portfolio}. Clearly, this immunizing portfolio is independent of $\Delta>0$ as the minmax problem \eqref{eq:VIZ} does not involve $\Delta$. Third, because the second order term in \eqref{eq:RobIm_error_bd} is proportional to $1+\norm{\theta}_1$, leverage can negatively affect the immunization performance.

\subsection{Relation to existing literature}\label{subsec:main_lit}

We relate Theorem \ref{thm:RobIm} to two questions in the immunization literature. First, when does exact sensitivity matching recover classical duration and factor methods and provide a global funding guarantee? Second, how does our local maxmin equity criterion differ from earlier maxmin and minimax formulations? The following corollary shows that when $I=J-1$ and there is no portfolio constraint beyond value matching, the immunizing portfolio can be solved explicitly.

\begin{cor}[Robust immunization with $I=J-1$]\label{cor:RobIm_exact}
Let everything be as in Proposition \ref{prop:minmax} and suppose that the only portfolio constraint is value matching \eqref{eq:value_match}, so the set of admissible portfolios is
\begin{equation}
    \cZ_0\coloneqq \set*{z\in \R^J: P(x)=\sum_{j=1}^J z_jP_j(x)}. \label{eq:cZ0}
\end{equation}
If $I=J-1$ and the square matrix $A_+$ in \eqref{eq:Abplus} is invertible, then the unique solution to \eqref{eq:VIZ} is $z^*=A_+^{-1}b_+$, with $V_I(\cZ)=0$.
\end{cor}

\begin{proof}
Immediate from the proof of Proposition \ref{prop:minmax}.
\end{proof}

\begin{rem}\label{rem:HD}
The special case of Corollary \ref{cor:RobIm_exact} with $I=J-1=1$ and $h_1(t)=t$ reduces to classical immunization that matches the bond value and duration \citep{FisherWeil1971,BierwagKhang1979}. To see this, recall that by the definition \eqref{eq:duration}, the duration of the cash flow $F$ equals the weighted average time to payment
\begin{equation*}
    D=\frac{\int_0^T t\e^{-ty(t)}\diff F(t)}{\int_0^T \e^{-ty(t)}\diff F(t)}.
\end{equation*}
Using the definition $x(t)=ty(t)$ and \eqref{eq:gateaux}, the duration can be rewritten as
\begin{equation*}
    D=\frac{\int_0^T t\e^{-x(t)}\diff F(t)}{\int_0^T \e^{-x(t)}\diff F(t)}=-\frac{P'(x)h_1}{P(x)}=b_1,
\end{equation*}
where $h_1(t)=t$ and we have used \eqref{eq:bi}. A similar calculation implies that the duration of the immunizing portfolio is
\begin{equation*}
    -\frac{\sum_{j=1}^J z_jP_j'(x)h_1}{\sum_{j=1}^J z_jP_j(x)}=-\frac{\sum_{j=1}^J z_jP_j'(x)h_1}{P(x)}=\sum_{j=1}^Ja_{1j}z_j
\end{equation*}
using value matching \eqref{eq:value_match} and \eqref{eq:aij}. Therefore if $z=A_+^{-1}b_+$, so $A_+z=b_+$, the duration is matched. By the same argument, setting $I=J-1$ and $h_i(t)=t^i$ reduces to \emph{high-order duration matching} ($I=J-1=2$ is convexity matching) \citep{ChambersCarletonMcEnally1988,NawalkhaLacey1988,PrismanShores1988}. If, instead, we use the basis functions corresponding to the factor loadings on the yield curve's level, slope, and curvature as described in Example \ref{exmp:pca}, then $A_+z = b_+$ amounts to factor duration matching \citep{LittermanScheinkman1991,Willner1996}. Thus, Corollary \ref{cor:RobIm_exact} nests these familiar methods as exactly identified cases of sensitivity matching.
\end{rem}

Exact sensitivity matching eliminates the effect of every yield curve change in the chosen span to first order. It does not generally prevent underfunding after a finite change. If the liability pays on a single date and the exact matching portfolio is nonnegative, convexity yields the following global result.

\begin{prop}[Guaranteed funding]\label{prop:maxmin_global}
Let everything be as in Corollary \ref{cor:RobIm_exact} and suppose that the liability pays out on a single date. If $z^*=A_+^{-1}b_+\ge 0$, then for all $h\in \spn\set{h_i}_{i=1}^I$ we have $E(z^*,x+h)\ge 0$.
\end{prop}

\begin{rem}\label{rem:maxmin}
When $I=J-1=1$ and $h_1(t)=t$, Proposition \ref{prop:maxmin_global} reduces to the global funding guarantee of \citet{FisherWeil1971}. With additional basis functions, the proposition extends the same guarantee to richer spans of yield curve changes, but it continues to require a single liability and a nonnegative exact matching portfolio. Earlier papers obtain global results for multiple liabilities under additional structure. \citet{BierwagKaufmanToevs1983} construct immunization strategies for multiple payments under one or more interest rate shocks. For the class of rate changes that he considers, \citet{shiu1988immunization} shows that global immunization requires the asset cash flows to be decomposable into portfolios that immunize the liability payments separately. Theorem \ref{thm:RobIm} does not impose this structure or claim a global guarantee for an arbitrary liability stream. Instead, it measures the best local protection available when exact matching may be impossible.

The distinction between immunization, maxmin choice, and minimax risk is important here. \citet{BierwagKhang1979} show that classical duration matching is a maxmin strategy under parallel shifts, and \citet{prisman1986immunization} extends the analysis to a setting with taxes. \citet{balbas1998can} distinguish an immunized portfolio, which attains a target in every admissible state, from a maxmin portfolio, which maximizes the worst payoff. \citet{Balbas2002} compute maxmin portfolios when immunization is infeasible. \citet{BalbasRomera2007} formulate maximin hedges for multiple liabilities under broad shock classes and practical portfolio constraints.

A separate strand measures risk with directional derivatives. \citet{bowden1997generalising} generalizes duration using directional derivatives, and \citet{BarberCopper1998} choose the portfolio that minimizes its maximum sensitivity over a shock set with infinitely many factors. Their minimax criterion is conceptually closest to $V_I(\cZ)$ in \eqref{eq:VIZ}. Our original problem \eqref{eq:maxmin_I} instead maximizes equity in the worst case. Theorem \ref{thm:RobIm} connects the two approaches: it shows that the maxmin equity problem reduces, to first order, to the minimax sensitivity problem and provides the corresponding error bound for finite stresses. Within this framework, we allow arbitrary deterministic liability cash flows, a number of stress directions greater than the number of bonds, and any closed portfolio constraint set. Proposition \ref{prop:RobIml2} and Section \ref{sec:jacobi} then provide a solution in closed form under an $\ell^2$ stress set and linear portfolio constraints, together with a representation in terms of exactly matching portfolios. Section \ref{sec:eval} evaluates these portfolios when short sale, leverage, and collateral constraints bind.
\end{rem}

\section{Extensions and implementation}\label{sec:extensions}

In this section, we discuss extensions of robust immunization such as using the $\ell^p$ norm and principal components and the numerical implementation.

\subsection{Robust immunization with \texorpdfstring{$\ell^p$}{} perturbations}\label{subsec:main_lp}
The portfolio solution in \eqref{eq:VIZ} can be obtained using linear programming techniques. However, computing the solution can become slow when the number of basis functions or payment dates is large, due to the large number of linear constraints. In addition, there is no closed-form expression for the portfolio, which makes it difficult to analyze how changes---such as adding basis functions---affect the solution.

To address these issues, we generalize the stress set, which leads to a portfolio solution that can be obtained in closed form. So far, we have defined the stress set as $\cH_I(\Delta)$ in \eqref{eq:cH_Delta}. More generally, to measure the magnitude of perturbations, for $p\in [1,\infty]$ we may use the $\ell^p$ norm on the payment dates $t=(t_1,\dots,t_N)$ defined by
\begin{equation}
    \norm{h(t)/t}_{p,t}\coloneqq \begin{cases}
        \left(\sum_{n=1}^N\abs{h(t_n)/t_n}^p\right)^{1/p}, & (p<\infty)\\
        \max_n\abs{h(t_n)/t_n}, & (p=\infty)
    \end{cases} \label{eq:lpnorm}
\end{equation}
and define the $\ell^p$ stress set by
\begin{equation}
    \cH_I^p(\Delta)\coloneqq \set*{h\in \spn\set{h_i}_{i=1}^I: \norm{h(t)/t}_{p,t}\le \Delta}. \label{eq:cHp_Delta}
\end{equation}
The case $p=\infty$ corresponds to what we treated in Section \ref{subsec:main_RobIm}. With this generalization, we need to redefine the minmax problem \eqref{eq:VIZ} using
\begin{align}
    \cW^p&\coloneqq \set{w\in \R^I: \norm{G'w}_p\le 1},\label{eq:cWp}\\
    V_I^p(\cZ) &\coloneqq \inf_{z\in \cZ}\sup_{w\in \cW^p}\seq{w,Az-b}, \label{eq:VIZp}
\end{align}
where $\norm{\cdot}_p$ denotes the usual $\ell^p$ norm for vectors. Noting the equivalence of norms for finite-dimensional spaces \citep[p.~13, Theorem 1.3]{TodaEME}, it is straightforward to generalize Theorem \ref{thm:RobIm} in this setting, though the expression for the high-order term in the error estimate in \eqref{eq:RobIm_error_bd} needs to be modified appropriately.

The special case of $p=2$ (Euclidean norm for perturbations) with linear constraints is particularly analytically tractable, as the following proposition shows.\footnote{In Section \ref{subsec:main_principal}, we show why it can sometimes be beneficial to impose multiple linear constraints in the portfolio solution.}

\begin{prop}[Robust immunization with $\ell^2$ perturbations]\label{prop:RobIml2}
Suppose Assumptions \ref{asmp:payouts}--\ref{asmp:H} hold and the portfolio constraint is given by $\cZ=\set{z\in \R^J:Rz=r}$, where $R\in \R^{M\times J}$ has full row rank and $r\in \R^M$. If $p=2$, $A$ has full column rank, and $\tilde{A}\coloneqq (GG')^{-1}A$, then the unique solution to the minmax problem \eqref{eq:VIZp} is
\begin{equation}
    z=(\tilde{A}'A)^{-1}\tilde{A}'b+(\tilde{A}'A)^{-1}R'[R(\tilde{A}'A)^{-1}R']^{-1}(r-R(\tilde{A}'A)^{-1}\tilde{A}'b), \label{eq:z_l2}
\end{equation}
where the inverses all exist.
\end{prop}

The portfolio solution can also be viewed as a projection of the liability sensitivity vector $b$ on the bond sensitivity matrix A. Specifically, consider the model
\begin{equation*}
b = Az + \varepsilon, \quad \text{where} \quad \varepsilon \mid A \sim \mathsf{N}\lro{0,G G'},
\end{equation*}
where $\mathsf{N}\lro{\mu,\Sigma}$ denotes the multivariate normal distribution with mean $\mu$ and variance-covariance matrix $\Sigma$. Using maximum likelihood estimation for the constrained problem yields the same solution as \eqref{eq:z_l2}. Hence, the maxmin portfolio \eqref{eq:z_l2}  corresponds to a constrained generalized least squares (GLS) solution.  

\subsection{Robust immunization as a regularized HD portfolio}\label{sec:jacobi}
As is well known, high-order duration matching ($I=J-1$, no portfolio constraint, and $h_i(t)=t^i$, which we refer to as \HD) need not perform well due to extreme leverage \citep{Mantilla-Garcia2022}. In the portfolio literature, a useful remedy to extreme portfolio weights is to apply shrinkage methods \citep{Ledoit2003,kozak2020shrinking}. Here we show that overidentification of the portfolio ($I > J-1$) can also be interpreted as a form of regularization, leading to less extreme portfolio weights and providing a theoretical explanation for why the robust immunization portfolio may outperform high-order duration matching.

It turns out that there is a close connection between high-order duration matching (Corollary \ref{cor:RobIm_exact}) and robust immunization with $\ell^2$ perturbations (Proposition \ref{prop:RobIml2}). To spell this out in more detail we introduce some terminology. For any matrix $A \in \R^{I\times J}$ with $\rank(A) = J$, define $A(s) \in \R^{J \times J}$ to be the matrix $A$ with row index in $s \subset \set{1,\dots,I}$ and $\abs{s} = J$. Similarly, let $b(s) \in \R^J$ denote the vector $b$ with row index in $s$. We call $z(s) = A(s)^{-1} b(s) \in \mathbb{R}^J$ an elemental estimate. \citet{jacobi1841formatione} showed that the ordinary least squares estimator can be expressed as the expected value of the elemental estimates, with the probability measure given by $\Pr(S = s) = \det\lr{A_+(s)}^2/\sum_s\det\lr{A_+(s)}^2$ (see also \citet{knight2018elemental}). In Appendix \ref{app:proof_cgls} we generalize these results to the constrained regression case. Applied to our context, the result is as follows.

\begin{prop}\label{prop:cgls}
Let $A_{\texttt{HD}} \in \R^{(J-1)\times J}$ be the bond sensitivity matrix corresponding to high-order duration matching. Let $A \in \R^{I \times J}$ ($I > J-1$) be the bond sensitivity matrix of the robust immunization portfolio in Proposition \ref{prop:RobIml2}, which can be partitioned as $A' = [A_{\texttt{HD}}', \ A(\set{J,\dots,I})']$, where $A(\set{J,\dots,I}) \in \R^{(I-J+1)\times J}$. Define $\tilde{A} \coloneqq (G G')^{-1} A$, and let
\begin{equation*}
A_+ = \begin{bmatrix}
a_0\\
A
\end{bmatrix}, \ 
\tilde{A}_+ = \begin{bmatrix}
a_0\\
\tilde{A}
\end{bmatrix}, \
b_+ = \begin{bmatrix}
1\\
b
\end{bmatrix} .
\end{equation*}
Then the solution to the minmax problem in \eqref{eq:z_l2} with a value matching constraint can be expressed as
\begin{equation}\label{eq:jacobi_cgls}
z = \frac{\sum_{1 \subset s} \det\lr{\tilde{A}_+(s)} \det\lr{A_+(s) } z(s) }{\sum_{1 \subset s} \det\lr{\tilde{A}_+(s)} \det\lr{A_+(s) }} \eqqcolon \sum_{[M] \subset s} \lambda(s) z(s),
\end{equation}
where $z(s) = A_+^{-1}(s) b_+(s) $, $\sum_{1 \subset s}$ denotes the sum over all subsets $s$ of cardinality $J$ that contain $1$, and
\begin{equation*}
\lambda(s) = \frac{\det\lr{\tilde{A}_+(s)} \det\lr{A_+(s) }}{\sum_{1 \subset s} \det\lr{\tilde{A}_+(s)} \det\lr{A_+(s) }} \quad  \text{with} \quad \sum_{1 \subset s} \lambda(s) = 1.
\end{equation*}
\end{prop}
One of the summands in \eqref{eq:jacobi_cgls} includes the set $s = \set{1,\dots,J}$. In that case, $z(s)$ is given by the \HD solution in Corollary \ref{cor:RobIm_exact}. The weight given to the \HD solution is proportional to $\det(A_+(s))$. Hence, if the sensitivity matrix of the \HD solution is close to singular, it receives little weight in the robust immunization portfolio. In this sense robust immunization with $I > J-1$ helps to regularize the portfolio solution, since extreme portfolio weights or leverage induced by a close to singular sensitivity matrix have less bearing on the portfolio solution in \eqref{eq:jacobi_cgls}.\footnote{A caveat is in order: the weights $\lambda(s)$ in \eqref{eq:jacobi_cgls} sum to one but need not be nonnegative, so \eqref{eq:jacobi_cgls} is in general an exact decomposition rather than a convex combination. Convexity holds, however, if the basis is orthonormalized on the payout dates, which is without loss of generality because the stress set \eqref{eq:cHp_Delta} depends on the basis only through its span (Proposition \ref{prop:invariance}). In that case $\tilde{A}=A$ and $\lambda(s)=\det\lr{A_+(s)}^2/\sum_{1\subset s}\det\lr{A_+(s)}^2\ge 0$.} 

Furthermore, the solution in \eqref{eq:jacobi_cgls} is a linear combination of different maxmin solutions, since by Proposition \ref{prop:minmax} and Theorem \ref{thm:RobIm}, $z(s)$ solves a minmax problem for every $s$ with $V_{\abs{s}}(\cZ) = 0$, where different basis functions  correspond to the rows $s$. This decomposition sheds light on the choice of basis functions. Consider the robust immunization portfolio with $I = J$, and compare this to the \HD solution. If the perturbation to the yield curve is solely due to the $I$th basis function, it follows from \eqref{eq:jacobi_cgls} that the robust immunization portfolio always does better than \HD, provided the weight given to the \HD portfolio is between zero and one. On the other hand, if the perturbation is a weighted average of several basis functions, the \HD portfolio may outperform robust immunization. The choice of basis functions thus hinges on how close they are to spanning the shock and on how much each basis function contributes to the shock. In Figure \ref{fig:R2} below, we empirically find that the first 10 basis functions strike a good balance.

\subsection{Robust immunization with principal components}\label{subsec:main_principal}

So far we have put no structure on the basis functions $\set{h_i}_{i=1}^I$ beyond Assumption \ref{asmp:H}. The stress set \eqref{eq:cHp_Delta} depends only on $\spn\set{h_i}_{i=1}^I$ (Proposition \ref{prop:invariance}) and the particular order or parameterization does not matter. However, in practice there may be a factor structure in the yield curve. For instance, a typical shift to the yield curve might be decomposed into the sum of a parallel shift and a nonparallel shift of a smaller size. In fact, according to Figure \ref{fig:R2} below, the first few basis functions explain a large fraction of the variation in the yield curve changes. Therefore it could be important to account for explanatory power of different basis functions in constructing the robust immunizing portfolio. In addition, a fund manager may believe that future short rates rise or fall uniformly across all horizons following, for example, a monetary policy shock, which primarily affects the level of the yield curve. In such a case, one would like to construct an optimal portfolio that is fully robust to level shocks, while allowing for exposure to other types of shocks that are deemed less likely.

We formalize this idea and extend Theorem \ref{thm:RobIm} and Proposition \ref{prop:RobIml2} to a setting in which the perturbation in a particular direction (principal component) could be larger. For any $\Delta_1,\Delta_2>0$, consider the following stress set:
\begin{multline}
    \cH_I^p(\Delta_1,\Delta_2)\\
    \coloneqq \set*{h\in \spn\set{h_i}_{i=1}^I:(\exists \alpha) \norm{\alpha h_1(t)/t}_{p,t}\le \Delta_1,\norm{h(t)/t-\alpha h_1(t)/t}_{p,t}\le \Delta_2}.\label{eq:cH_Delta1}
\end{multline}
Choosing $h\in \cH_I^p(\Delta_1,\Delta_2)$ amounts to perturbing the yield curve in the direction spanned by the first component ($h_1(t)/t$) by a magnitude at most $\Delta_1$, and then perturbing in an arbitrary direction spanned by the first $I$ basis functions by a magnitude at most $\Delta_2$. Thus setting $\Delta_1\gg \Delta_2$ captures the idea that $h_1$ is the first principal component. In this setting, we can generalize Theorem \ref{thm:RobIm} and Proposition \ref{prop:RobIml2} as follows.

\begin{thm}[Robust immunization with principal components]\label{thm:RobIm1}
Let the assumptions of Proposition \ref{prop:minmax} hold and suppose the set
\begin{equation}
    \cZ_1\coloneqq\set*{z\in \cZ:\sum_{j=1}^J a_{1j}z_j=b_1}\label{eq:cZ1}
\end{equation}
is nonempty, where $a_{1j}$ and $b_1$ are defined by \eqref{eq:aij} and \eqref{eq:bi} with $i=1$. Let $\cH_I^p(\Delta_1,\Delta_2)$ be as in \eqref{eq:cH_Delta1}. Then the following statements are true.
\begin{enumerate}
    \item The guaranteed equity satisfies
    \begin{equation}
        \lim \frac{1}{\Delta_2}\sup_{z\in \cZ_1}\inf_{h\in \cH_I^p(\Delta_1,\Delta_2)} E(z,x+h)=-P(x) V_I^p(\cZ_1),\label{eq:maxmin1}
    \end{equation}
    where the limit is taken over $\Delta_1,\Delta_2\to 0$, $\Delta_1/\Delta_2\to\infty$, and $\Delta_1^2/\Delta_2\to 0$.
    \item Letting $z^*\in \cZ_1$ be the solution to the minmax problem \eqref{eq:VIZ} with portfolio constraint $\cZ_1$, we have
    \begin{equation}
        \sup_{h\in \cH_I(\Delta_1,\Delta_2)}\abs{E(z^*,x+h)}\le \Delta_2 P(x)\left(V_I(\cZ_1)+O(\Delta_2+\Delta_1^2/\Delta_2)\right).\label{eq:RobIm_error_bd1}
    \end{equation}
    \label{item:thmPCA2}
\end{enumerate}
\end{thm}

Note that part \ref{item:thmPCA2} uses the $\ell^\infty$ norm, but can be extended  to any other $\ell^p$ norm for $p \ge 1$. Imposing the portfolio constraint $\cZ_1\subset \cZ$ may improve or worsen the performance. To explain why, we first present the following simple result.

\begin{prop}[Monotonicity of minmax value]\label{prop:monotoneVIZ}
Let everything be as in Theorem \ref{thm:RobIm1}. If $I<I'$ and $\cZ\subset \cZ'$, then $V_I^p(\cZ)\le V_{I'}^p(\cZ)$ and $V_I^p(\cZ)\ge V_I^p(\cZ')$.
\end{prop}

The claim $V_I^p(\cZ)\le V_{I'}^p(\cZ)$ is obvious because the more basis functions we include, the more freedom nature has to select adversarial perturbations. The claim $V_I^p(\cZ)\ge V_I^p(\cZ')$ is also obvious because the larger the set of admissible portfolios is, the more freedom the fund manager has to select portfolios.

Comparing to \eqref{eq:cH_Delta1} with \eqref{eq:cHp_Delta} and applying the triangle inequality
\begin{equation*}
    \norm{h(t)/t}_{p,t}\le \underbrace{\norm{\alpha h_1(t)/t}_{p,t}}_{\le \Delta_1}+\underbrace{\norm{h(t)/t-\alpha h_1(t)/t}_{p,t}}_{\le \Delta_2},
\end{equation*}
we obtain $\cH_I^p(\Delta_1,\Delta_2)\subset \cH_I^p(\Delta_1+\Delta_2)$. Therefore, to first order, the maximum portfolio return loss can be bounded as
\begin{equation*}
    \underbrace{\Delta_2 V_I^p(\cZ_1)}_\text{Theorem \ref{thm:RobIm1}}\le (\Delta_1+\Delta_2)V_I^p(\cZ_1)\ge \underbrace{(\Delta_1+\Delta_2)V_I^p(\cZ)}_\text{Theorem \ref{thm:RobIm}},
\end{equation*}
where the right inequality follows from Proposition \ref{prop:monotoneVIZ}. Thus if $\Delta_1\gg \Delta_2$ in typical situations (which we document in Figure \ref{fig:R2}), then imposing the constraint $\cZ_1$ in \eqref{eq:cZ1} improves the performance because the loss in the minmax value $V_I^p(\cZ_1)\ge V_I^p(\cZ)$ from imposing the constraint is compensated by the gain in the coefficient $\Delta_2\ll \Delta_1+\Delta_2$. However, if it so happens that $\Delta_1\sim \Delta_2$, then imposing the constraint $\cZ_1$ worsens the performance.

This result sheds further light on the sometimes poor performance of high-order duration matching ($I=J-1$, no portfolio constraint, and $h_i(t)=t^i$). First, as we increase $I$ while setting $I=J-1$, both the span of basis functions and the set of admissible portfolios $\cZ$ expand. Because increasing $I$ makes $V_I^p(\cZ)$ larger but expanding $\cZ$ makes it smaller, the combined effect could go either way. This observation explains the poor performance of high-order duration matching. Second, in the setting of Theorem \ref{thm:RobIm1}, if $\Delta_1\sim \Delta_2$, then imposing the constraint $\cZ_1$ worsens the performance. To see why, by Proposition \ref{prop:monotoneVIZ} we have $V_I^p(\cZ_1)\ge V_I^p(\cZ)$, so if $\Delta_1\sim \Delta_2$, then
\begin{equation*}
    \underbrace{\Delta_2 V_I^p(\cZ_1)}_\text{Theorem \ref{thm:RobIm1}}> \underbrace{(\Delta_1+\Delta_2)V_I^p(\cZ)}_\text{Theorem \ref{thm:RobIm}}.
\end{equation*}

\begin{rem}
Theorem \ref{thm:RobIm1} can be further generalized if we allow larger perturbations spanned by the first few basis functions. For instance, if we use the first two basis functions, we can define $\cH_I^p(\Delta_1,\Delta_2,\Delta_3)$ analogously to \eqref{eq:cH_Delta1} by incorporating the constraints $\norm{\alpha_ih_i(t)/t}_{p,t}\le \Delta_i$ for $i=1,2$ and 
\begin{equation*}
\norm{h(t)/t-\alpha_1h_1(t)/t-\alpha_2h_2(t)/t}_{p,t}\le \Delta_3.    
\end{equation*}
The portfolio constraint \eqref{eq:cZ1} then becomes
\begin{equation}
    \cZ_2\coloneqq\set*{z\in \cZ:\sum_{j=1}^J a_{ij}z_j=b_i~\text{for}~i=1,2},\label{eq:cZ2}
\end{equation}
and the maxmin formula \eqref{eq:maxmin1} involves $V_I^p(\cZ_2)$. By a similar argument as above, imposing the constraint $\cZ_2$ improves the performance relative to $\cZ_1$ if $\Delta_2\gg \Delta_3$, whereas it worsens the performance if $\Delta_2\sim \Delta_3$.
\end{rem}

\subsection{Implementation}\label{subsec:main_implement}

To implement robust immunization, we need to choose the basis functions $\set{h_i}_{i=1}^I$. Although the conclusion of Theorem \ref{thm:RobIm} holds regardless of the choice of the basis functions, here we propose a particular choice.

\begin{figure}[!htb]
\centering
\begin{subfigure}{0.48\linewidth}
    \includegraphics[width=\linewidth]{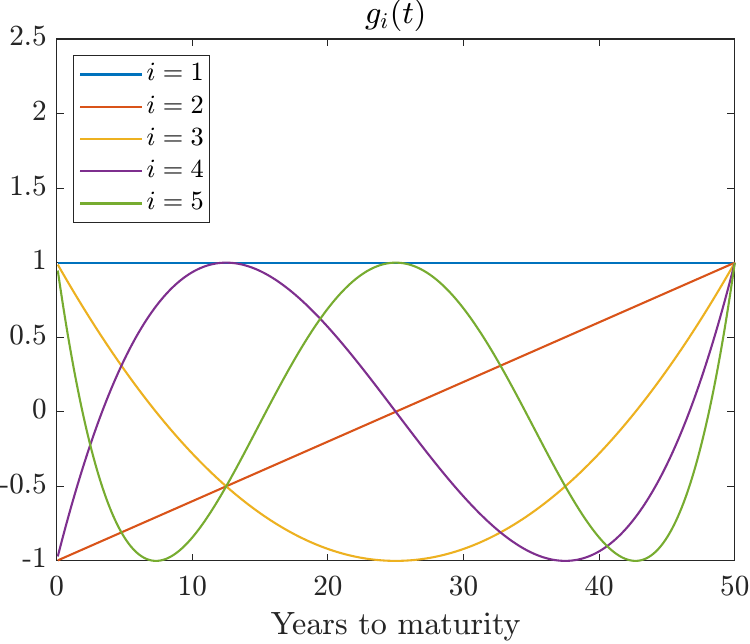}
\caption{Yield curve.}\label{fig:basis_g}
\end{subfigure}
\begin{subfigure}{0.48\linewidth}
    \includegraphics[width=\linewidth]{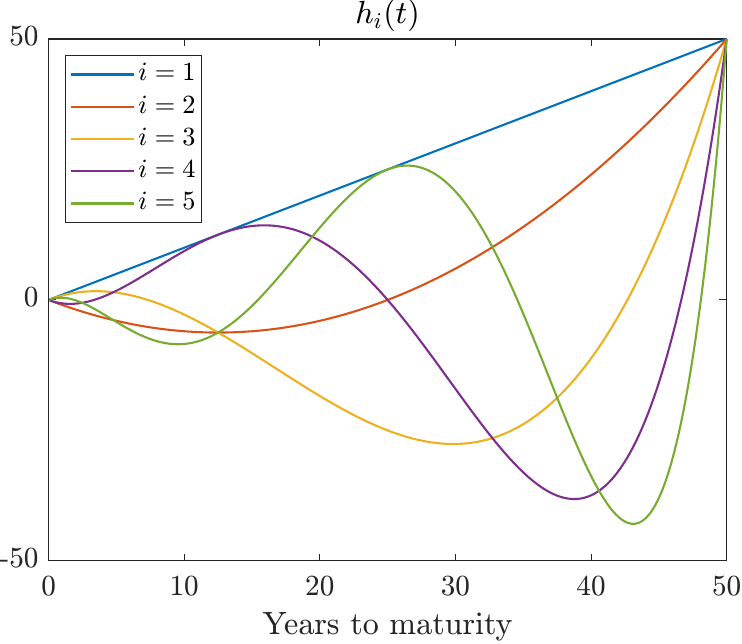}
\caption{Cumulative discount rate.}\label{fig:basis_h}
\end{subfigure}
\caption{Basis functions of robust immunization.}\label{fig:basis}
\end{figure}

For each $i$, it is natural to choose $h_i$ such that $h_i$ is a polynomial of degree $i$ with $h_i(0)=0$, for Assumption \ref{asmp:H} then holds (Lemma \ref{lem:poly}). By basis invariance (Proposition \ref{prop:invariance}), any choice of such a basis will result in the same immunizing portfolio. However, we suggest using Chebyshev polynomials because they enjoy desirable numerical properties such as the ability to approximate continuous functions \citep[Ch.~2--4]{Trefethen2019}. To be more specific, let $T_n:[-1,1]\to \R$ be the degree-$n$ Chebyshev polynomial defined by $T_n(\cos\theta)=\cos n\theta$ and setting $x=\cos\theta$. We map $[0,T]$ to $[-1,1]$ using the affine transformation $t\mapsto x=2t/T-1$, and define $g_i:[0,T]\to \R$ by
\begin{equation}
    g_i(t)=T_{i-1}(2t/T-1) \label{eq:gi}
\end{equation}
so that we can allow any (continuous) perturbation to the yield curve for $t\in [0,T]$. Then we can define the basis functions for perturbing the cumulative discount rate by $h_i(t) = t g_i(t)$. Figure \ref{fig:basis_g} shows the graphs of $g_i$ in \eqref{eq:gi} for a maturity $T=50$ years, which are the rows of the matrix $G$ in Proposition \ref{prop:minmax}. Figure \ref{fig:basis_h} shows the graphs of the basis functions $h_i(t)$. The Chebyshev basis is also economically interpretable: the first three yield curve loadings $g_1$, $g_2$, $g_3$ are constant, linear, and quadratic in maturity, and thus mimic the level, slope, and curvature factors that dominate variation in the yield curve \citep{LittermanScheinkman1991} (see also Example \ref{exmp:pca}).

Because Theorem \ref{thm:RobIm} takes the number of basis functions $I$ as given, a natural question is how to select it. Choosing a small $I$ restricts the stress set and may lead to lack of robustness against model misspecification. To address this concern, we evaluate the goodness-of-fit of approximating discount rate changes by basis functions. For this purpose, we use the daily yield curve data described in Section \ref{sec:eval}. Let $I$ be the number of basis functions to include, $d$ the number of days ahead, and $\set{t_n}_{n=1}^N$ the set of terms (in years) to evaluate the cumulative discount rates, where we set $t_n=n/12$ and $N=360$, which corresponds to a 30-year horizon with monthly payments. Let $y_s(t)$ be the yield curve on day $s$ at maturity $t$. We use the following procedure.

\begin{enumerate}
    \item For each day $s$ and term $t_n$, calculate the $d$-day-ahead change in the yield curve $y_{s+d}(t_n) - y_s(t_n)$.
    \item For each $s$ and $d$, estimate
    \begin{equation}
        y_{s+d}(t_n) - y_s(t_n) = \sum_{i=1}^I \gamma_{isd}g_i(t_n) + \epsilon_{sd}(t_n), \quad n=1,\dots,N \label{eq:OLS}
    \end{equation}
    by ordinary least squares (OLS), where $g_i$ is as in \eqref{eq:gi}. 
    
    \item For each basis function $i$, decompose the $R^2$ of the regression  \eqref{eq:OLS} using the Shapley value (see \citet{huettner2012axiomatic}):\footnote{The Shapley value of the $R^2$ has some desirable properties, such as efficiency and monotonicty \citep{huettner2012axiomatic}. Furthermore, the order of the regressors is irrelevant.} 
    \begin{equation*}
    R_i^2 = \frac{1}{I} \sum_{S \setminus \set{i}} \binom{I-1}{\abs{S}}^{-1} \lro{ R^2(S \cup \set{i}) - R^2(S)}.
    \end{equation*}
    The sum is taken over all subsets $S$ of $\set{1,\dots,I}$ that do not include basis function $g_i$. To calculate $R^2(S)$ for a model that includes basis functions in $S$, we use $0$ as the benchmark instead of the sample mean, since $g_1 \equiv 1$ is already a constant function. As a result, $R_i^2$ measures the explanatory power of basis function $i$ relative to the other basis functions.

    \item Let $\set{\hat{\gamma}_{isd}}_{i=1}^I$ be the OLS estimator, calculate the overall goodness-of-fit measure 
    \begin{equation}
        R_{d}^2 \coloneqq \frac{\sum_{s=1}^S\sum_{n=1}^N \left(\sum_{i=1}^I \hat{\gamma}_{isd}g_i(t_n) \right)^2  }{\sum_{s=1}^S\sum_{n=1}^N (y_{s+d}(t_n) - y_s(t_n))^2}.\label{eq:R2}
    \end{equation}
\end{enumerate}

\begin{figure}[!htb]
\centering
\includegraphics[width=0.48\linewidth]{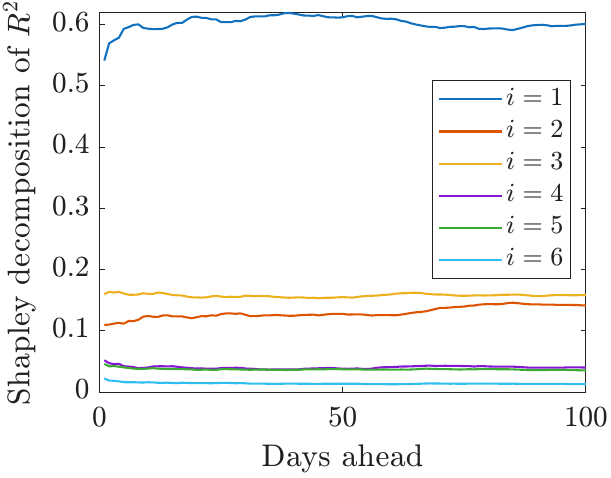}
\hfill 
\includegraphics[width=0.48\linewidth]{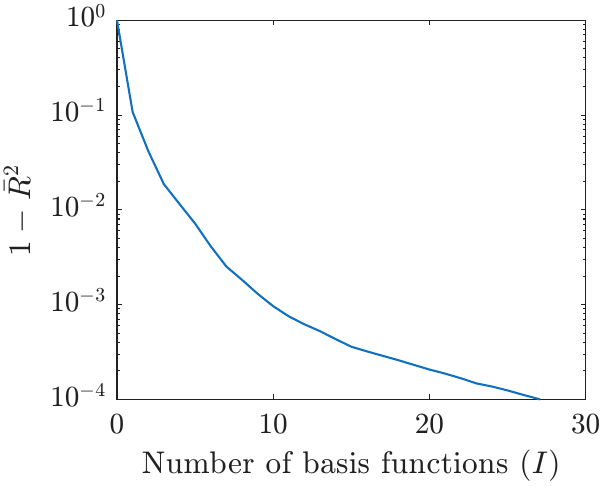}
\caption{Goodness-of-fit of yield curve change approximation.}\label{fig:R2}
\caption*{\footnotesize Note: The left panel shows the decomposed $R^2$ using the Shapley value  corresponding to regression \eqref{eq:OLS} with $I=6$ basis functions. The Shapley values are averaged across all dates in the training window. The right panel shows the average $1-R_{d}^2$ across $d$ as we increase the number of basis functions $I$. See Section \ref{sec:eval} for data description.}
\end{figure}

To avoid selecting the number of basis functions $I$ in sample, we tune it on yield curve data from November 1985 to December 1995 only, and evaluate the hedging simulations reported below on the holdout period from January 1996 to December 2023.\footnote{All results are very similar when using the entire sample.} The left panel of Figure \ref{fig:R2} shows the Shapley decomposition of the $R^2$ with $I=6$ basis functions. The Shapley values are averaged across all dates in the training window. The explanatory power of each basis function is roughly constant across different horizons $d$. The first basis function (constant) explains around 60\% of variations in the yield curve changes. In more than 98\% of the periods, the $R^2$ is above 95\%.\footnote{Appendix \ref{app:kernel_basis} repeats this experiment using the kernel basis functions of \citet{filipovic2025stripping}. We find that the Chebyshev basis has higher explanatory power.} Furthermore, most of the explanatory power is contributed by the first, second and third basis functions, while the other basis functions generally contribute less than 5\%. The right panel shows the unexplained component $1-R_{d}^2$ averaged across $d$ as we include more basis functions. We can see that setting $I=10$ captures about 99.88\% ($1-R_{d}^2\sim 10^{-3}$) of variations in the yield curve changes.

We now describe the algorithm to implement robust immunization in practice. Although the underlying theory may not be familiar to practitioners, the implementation requires little more than basic linear algebra and linear programming.

\begin{framed}
\begin{oneshot}[Robust Immunization]
\quad
\begin{enumerate}
    \item Let $\mathbf{t}=(t_1,\dots,t_N)$ be the $1\times N$ vector of asset/liability payout dates and $T=t_N$ be the planning horizon. Let $\mathbf{y}=(y_1,\dots,y_N)$ be the $1\times N$ vector of yields, $\mathbf{f}=(f_1,\dots,f_N)$ the $1\times N$ vector of liabilities, and $\mathbf{F}=(f_{jn})$ the $J\times N$ matrix of bond payouts.
    \item Let $I\ge J-1$, define the basis functions by \eqref{eq:gi}, evaluate at each $t_n$, and construct the $I\times N$ matrix of basis functions $\mathbf{H}=(t_n g_i(t_n)) = (h_i(t_n))$ and $\mathbf{G}=(g_i(t_n))$. Define the $1\times N$ vector of zero-coupon bond prices $\mathbf{p}=\exp(-\mathbf{y}\odot \mathbf{t})$, where $\odot$ denotes entry-wise multiplication (Hadamard product).
    \item Define the $I\times J$ matrix $A$, $I\times 1$ vector $b$, and $1\times J$ vector $a_0$ by
    \begin{align*}
        A&\coloneqq (\mathbf{H}\diag(\mathbf{p})\mathbf{F}')/(\mathbf{p}\mathbf{f}'),&
        b&\coloneqq \mathbf{H}\diag(\mathbf{p})\mathbf{f}'/(\mathbf{p}\mathbf{f}'),&
        a_0&\coloneqq \mathbf{p}\mathbf{F}'/(\mathbf{p}\mathbf{f}'),
    \end{align*}
    where $\diag(\mathbf{p})$ denotes the diagonal matrix with diagonal entries given by $\mathbf{p}$. Define the $(I+1)\times J$ matrix $A_+$ and $(I+1)\times 1$ vector $b_+$ by
    \begin{equation*}
        A_+\coloneqq \begin{bmatrix}
        a_0 \\ A
        \end{bmatrix} \quad \text{and} \quad 
        b_+\coloneqq \begin{bmatrix}
        1 \\ b
        \end{bmatrix}.
    \end{equation*}
    \item If $I=J-1$ and there are no portfolio constraints, calculate the immunizing portfolio as $z^*=A_+^{-1}b_+$. Otherwise, numerically solve the minmax problem \eqref{eq:VIZ} (if using the $\ell^\infty$ norm) or use \eqref{eq:z_l2} (if using the $\ell^2$ norm and constraints are linear). Convex inequality constraints, such as short sale restrictions, can be easily incorporated. The resulting optimization problem can be solved using quadratic programming.
\end{enumerate}
\end{oneshot}
\end{framed}

Note that the inner maximization in \eqref{eq:VIZ} is a linear programming problem with $I$ variables and $2N$ inequality constraints, which is straightforward to solve numerically even when $N$ is large (a few hundred in typical applications). The outer minimization is a convex minimization problem with $J$ variables, which is also straightforward to solve numerically. However, in simulations in Section \ref{sec:eval}, repeatedly solving the linear programming problem can be computationally expensive. That is why we only consider the $\ell^2$ solution in the simulations below.

\section{Evaluation: static and dynamic hedging}\label{sec:eval}

In this section, we evaluate the performance of robust immunization and other existing methods using a numerical experiment in static and dynamic settings.

\subsection{Data and yield curve model}\label{subsec:eval_data}

We obtain daily U.S. Treasury nominal yield curve data from November 25, 1985 to December 2023 from \citet{liu2021reconstructing}.\footnote{\url{https://sites.google.com/view/jingcynthiawu/yield-data}. The estimated yields of \citet{liu2021reconstructing} go back all the way to 1961, but we only use their data beyond 11/25/1985 when bonds with a maturity of 30 years were introduced in the market.} These yield curves are estimated using a nonparametric method that accommodates general yield curve perturbations. This is important because more complex perturbations require a greater number of basis functions, which our approach can handle, unlike high-order duration matching.  We use the initial window from November 1985 to December 1995 to set the number of basis functions to $I = 10$  (Section~\ref{subsec:main_implement}), and evaluate hedging performance out of sample on the remaining dates. This leaves $S = 6{,}901$ evaluation dates spanning January 1996 to December 2023.

% \begin{rem}\label{rem:extrapolate}

% \citet{gurkaynak2007us} caution against extrapolation of the yield curve beyond the maximum available bond maturity. Anticipating our empirical application, we need to obtain yields with maturity up to 50 years. Since extrapolation is still necessary in this case, we extrapolate the forward rate by a constant beyond the 30-year maturity. This approach is motivated by no-arbitrage arguments which stipulate that the long term forward rate is constant \citep{dybvig1996long}. In Appendix \ref{app:bias}, we show how the constant forward rate assumption affects our estimate of the yield curve.
% \end{rem}

The observed yield-curve data provide only one realized path and thus inadequate for evaluating the performance in a dynamic hedging experiment. Therefore in addition to the observed yield curve data, we also use simulated yield curves generated from the term structure model of \citet{longstaff1992interest}.\footnote{An earlier version of our paper used the regime-switching model of \citet{AngBekaertWei2008} to simulate the yield curve. All results are robust to this modeling choice.} By simulating yields from this two-factor equilibrium model, we can evaluate the performance of various immunization methods under a wide variety of yield curves. See Appendix \ref{sec:LS} for details.

\subsection{Cash flow and immunization methods}\label{subsec:eval_method}
We consider five cash flow schemes for the liability, assuming a fixed time horizon of $T=30$ years. The specifications are as follows: (i) a constant cash flow of 1 each month throughout the horizon (\texttt{fullHorizon}); (ii) a constant cash flow of 1 after year 20 (\texttt{longRun}); (iii) a constant cash flow of 1 between years 10 and 20 (\texttt{medium}); (iv) a cash flow of 1 between years 0 and 10, and again between years 20 and 30 (\texttt{shortAndLong}); and (v) a monthly cash flow proportional to the survival probability of a cohort of retirees aged 65 under a Gompertz mortality law (\texttt{pension}), representing a defined-benefit pension in payment.\footnote{We set the modal age at death to 86 years and the dispersion parameter to 9.5 years, standard values in the actuarial literature.} In all cases, we normalize the cash flows so that their cumulative sum equals 1. The bonds available for trade are zero-coupon bonds with face value 1 and maturities of 1, 2, 5, 10, and 20 years. We intentionally choose a long maturity of 30 years for the cash flows because it is of interest to study how the yield curve at the long end affects the performance of the immunization methods; the \texttt{longRun} and \texttt{pension} liabilities in particular resemble the long-dated obligations of pension funds and life insurers.

To make the experiment more realistic, we consider several different frictions that constrain leveraged liability hedgers in practice. Because Theorem~\ref{thm:RobIm} holds for any closed constraint set $\cZ$ (Assumption~\ref{asmp:Z}), the theoretical formulation accommodates these frictions without modification. We impose three: no short sales ($\theta_j\ge 0$, equivalently $\norm{\theta}_1=1$); a gross-leverage cap $\norm{\theta}_1\le L=2$; and a collateral budget $\sum_{j=1}^J h_j\abs{\theta_j}\le B=0.05$, with repo haircuts $h_j$ rising from $0.5\%$ at the short end to $5\%$ at twenty years, in line with the Basel III supervisory haircuts for sovereign collateral \citep{BCBS2017} and with the leverage and collateral buffers of liability-driven investment mandates \citep{BoE2022FSR}. These constraints capture important practical limits: when rates rise, collateral calls on interest rate hedges can force pension funds to sell government bonds \citep{Breeden2022,Pinter2023,JansenKlinglerRanaldoDuijm2026}.

Under these constraints, we consider three immunization methods. The first method is our proposed robust immunization method (\RI) with the Chebyshev polynomial basis for the yield curve in \eqref{eq:gi} with $T = 30$. Given the right panel of Figure \ref{fig:R2}, we set the number of basis functions to $I=10$. Motivated by Theorem \ref{thm:RobIm1} and the left panel of Figure \ref{fig:R2}, we also impose additional matching constraints: value matching only ($\cZ_0$ in \eqref{eq:cZ0}), value- and duration matching ($\cZ_1$ in \eqref{eq:cZ1}), and value-, duration- and convexity matching. We denote these methods by \RI(0), \RI(1), \RI(2). Throughout the simulation, we report results only for the robust immunization portfolio based on the $\ell^2$ norm (see \eqref{eq:z_l2}), as it is computationally much faster than the portfolio based on the $\ell^\infty$ norm (see \eqref{eq:VIZ}).\footnote{Appendix \ref{app:eval_linf} shows that using the $\ell^\infty$ norm instead produces similar results.} 

The second method is key rate duration matching (\KRD) proposed by \citet{Ho1992} and explained in Appendix \ref{sec:KRD}. In short, this method is designed to match the asset and liability sensitivities to interest rate changes at prespecified maturities. 

The third method builds on high-order duration matching (\HD) explained in Remark \ref{rem:HD}, which is a special case of robust immunization by setting $I=J-1$ and $h_i(t)=t^i$. Exact high-order duration matching cannot be formed under any of the frictions we consider: the exactly identified matching problem has a unique solution whose gross leverage violates every constraint, the empirical counterpart to Remark~\ref{rem:maxmin}. Any feasible version must therefore relax exact matching. Robust immunization relaxes it with the minmax objective, which by Theorem~\ref{thm:RobIm} remains a robust immunizing portfolio under the constraint. As the feasible benchmark, we define least-squares high-order duration matching (\LSHD) as follows. Let $A_{\texttt{HD}}\in\R^{(J-1)\times J}$ and $b\in\R^{J-1}$ be the sensitivity matrix and vector defined by \eqref{eq:aij} and \eqref{eq:bi} with $h_i(t)=t^i$. The \LSHD portfolio solves
\begin{equation}
\min_{z\in\cZ}\ \sum_{i=1}^{J-1}\lr{\frac{(A_{\texttt{HD}}z)_i-b_i}{b_i}}^2, \label{eq:LSHD}
\end{equation}
where $\cZ$ imposes value matching \eqref{eq:value_match} together with the friction under consideration, so the objective minimizes the equally weighted squared relative mismatch of each moment.\footnote{Exact matching is invariant to how the moments are scaled, so any least-squares version must choose a weighting; we weight relative mismatches equally, so that a one-percent error in convexity counts as much as a one-percent error in duration. The minmax objective requires no such choice. Note that the relative mismatches are well defined since $b_i>0$ for all $i$.} Without frictions, the least-squares problem is exactly identified, the minimum in \eqref{eq:LSHD} is zero, and \LSHD coincides with \HD.

% The third method builds on high-order duration matching (\HD) explained in Remark \ref{rem:HD}, which is a special case of robust immunization by setting $I=J-1$ and $h_i(t)=t^i$. Exact high-order duration matching cannot be formed under any of the frictions we consider: the exactly identified matching problem has a unique solution whose gross leverage violates every constraint, the empirical counterpart to Remark~\ref{rem:maxmin}. Any feasible version must therefore relax exact matching. Robust immunization relaxes it with the minmax objective, which by Theorem~\ref{thm:RobIm} remains a robust immunizing portfolio under the constraint; as the feasible benchmark, we define least-squares high-order duration matching (\LSHD), which imposes the constraint exactly and matches the $J-1$ moments in least squares, minimizing the equally weighted squared relative mismatch of each moment.\footnote{Exact matching is invariant to how the moments are scaled, so any least-squares version must choose a weighting; we weight relative mismatches equally, so that a one-percent error in convexity counts as much as a one-percent error in duration. The minmax objective requires no such choice.} Unconstrained, the least-squares problem is exactly identified and \LSHD coincides with \HD.

\subsection{Static hedging}\label{subsec:eval_constr}
Suppose that on date $s$, the fund manager immunizes future cash flows with a bond portfolio $z_s = (z_{sj})$ constructed by any of the immunization methods. Letting $x_s$ be the cumulative discount rate on date $s$ and $h$ be a perturbation, we evaluate the performance of each method using the \emph{funding ratio} defined by
\begin{equation}
    \varphi_s(h)\coloneqq \frac{1}{P(x_s+h)}\sum_{j=1}^J z_{sj}P_j(x_s+h). \label{eq:funding_ratio}
\end{equation}
We suppose that the fund manager is worried about underfunding, so we further define the \emph{underfunding ratio} by
\begin{equation}
    1-\min\set{\varphi_s(h),1}. \label{eq:ufr}
\end{equation}
A higher underfunding ratio indicates a larger funding shortfall. As we are interested in realistic yield changes and portfolio holding periods, we take the perturbation $h$ to be the realized change in the cumulative discount rate from date $s$ to date $s+d$, for holding periods of
$d=1,\dots,100$ trading days and formation dates $s$ spanning 1996--2023.
Note the static nature of this experiment: perturbations are one-shot and
discount rates adjust instantaneously, so the maturities of the bonds held do
not shorten.

Table~\ref{t:constr} reports the $95$th percentile of the underfunding ratio \eqref{eq:ufr}. Under the leverage cap (Panel B), \RI(2) is feasible for four of the five liabilities and outperforms \KRD in each, by roughly fifteenfold for \texttt{pension} ($0.15\%$ against $2.27\%$) and ninefold for \texttt{medium}. It also outperforms \LSHD in all but \texttt{medium} ($0.60\%$ against $0.50\%$), where spending the full leverage budget on approximating the third and fourth moments pays for a liability near the boundary of attainable convexity. We observe that the performance of \RI(2) is stable across liabilities: \RI(2) stays between $0.15\%$ and $0.82\%$, whereas \LSHD ranges up to $1.06\%$ and \KRD from $0.96\%$ to $5.29\%$, faring worst for the \texttt{medium} liability that key rates cannot hedge without convexity control. 
Robust immunization is reliably hedged whatever the liability's shape, while the benchmarks are not.

The feasibility pattern in Table~\ref{t:constr} reflects the leverage that each method requires. Exact \HD demands a gross leverage of $43$ for \texttt{fullHorizon} and above $200$ for \texttt{longRun}, so it is infeasible in every panel and omitted, whereas the unconstrained \RI\ portfolios require a median gross leverage of only $1.1$ (\RI(0)), $2.3$ (\RI(1)), and $2.6$ (\RI(2)). Section~\ref{sec:jacobi} explains this ordering: \RI\ portfolios are constructed as averages over different \HD portfolios, which naturally reduces leverage, and imposing more matching constraints implies that the average is taken over fewer \HD portfolios, which tends to increase leverage.

The performance ranking in Table~\ref{t:constr} is also stable across constraints: the collateral budget (Panel C) tracks the leverage cap throughout, and the no-short-sale constraint (Panel A) is the most restrictive, since convexity matching requires short sales for \texttt{medium} and the long-dated liabilities. The constraints also frequently improve out-of-sample performance: the no-short-sale restriction lowers \RI(1)'s underfunding from $1.05\%$ to $0.42\%$ for \texttt{medium}, and lowers \LSHD, whose unconstrained solution is exact \HD, from $4.68\%$ to $0.71\%$ for \texttt{fullHorizon}. This is a familiar form of regularization since constraints act as shrinkage on the portfolio weights, cutting the large offsetting positions that make an unconstrained solution sensitive to model error \citep{JagannathanMa2003}.

The one exception is \texttt{longRun}, whose twenty-four-year duration exceeds the longest bond. Convexity matching requires a gross leverage near eleven, so \RI(2) is infeasible under every constraint. Under the leverage cap robust immunization falls back to duration matching, \RI(1), and a feasible \KRD hedge, which matches duration at lower leverage, does slightly better ($1.74\%$ against $2.27\%$, with \LSHD at $2.48\%$). Without short sales even duration matching is unattainable, and the gap widens: \RI(0) records $6.90\%$ against $2.12\%$ for the benchmarks, which collapse onto the same corner portfolio concentrated in the twenty-year bond. Every long-only portfolio here carries an irreducible duration gap, so the cell measures how a method allocates an unavoidable failure: the benchmarks minimize the duration gap, while \RI(0) minimizes the worst case over all curve shapes and accepts a larger duration gap in exchange for protection against non-parallel shifts. Level-dominated realized variation rewards the former ex post; the ranking reverses under the worst-case criterion that \RI(0) optimizes. This is the regime in which the constraint eliminates $\RI$'s main advantage---the ability to impose higher-order matching.

\begin{figure}[!htb]
\centering
\begin{subfigure}{0.48\linewidth}
    \includegraphics[width=\linewidth]{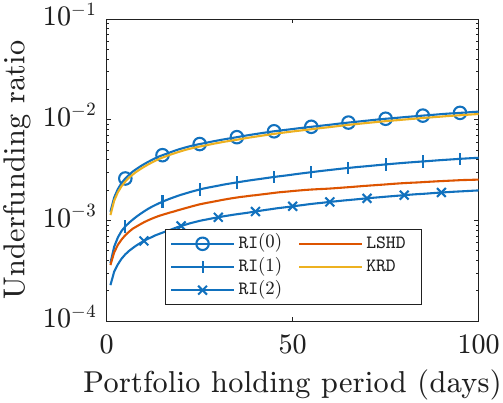}
\caption{\texttt{fullHorizon}.}\label{fig:ufr_fullHorizon}
\end{subfigure}
\begin{subfigure}{0.48\linewidth}
    \includegraphics[width=\linewidth]{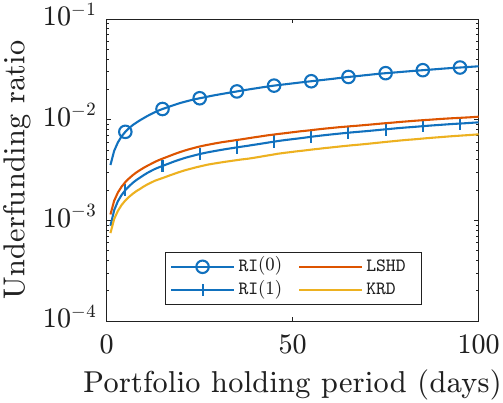}
\caption{\texttt{longRun}.}\label{fig:ufr_longRun}
\end{subfigure}\\
\begin{subfigure}{0.48\linewidth}
    \includegraphics[width=\linewidth]{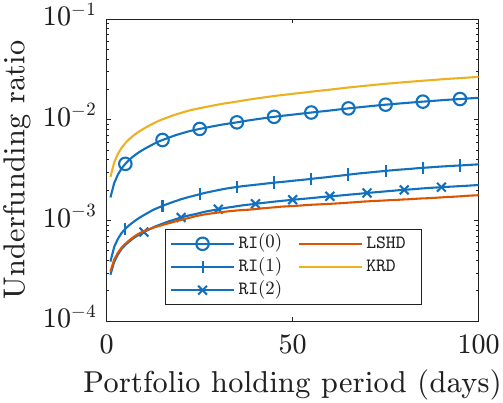}
\caption{\texttt{medium}.}\label{fig:ufr_medium}
\end{subfigure}
\begin{subfigure}{0.48\linewidth}
    \includegraphics[width=\linewidth]{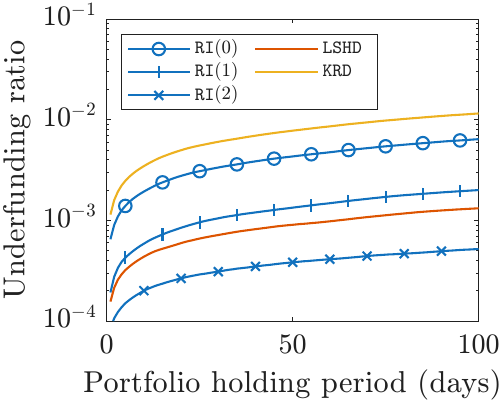}
\caption{\texttt{pension}.}\label{fig:ufr_pension}
\end{subfigure}
\caption{Underfunding ratio for different holding periods under the leverage cap.}\label{fig:ufr_horizon}
\caption*{\footnotesize Note: Average underfunding ratio \eqref{eq:ufr} over the 1996--2023 holdout as a function of the holding period, for the leverage-capped ($\norm{\theta}_1\le2$) version of each method. \RI(2) is infeasible for \texttt{longRun} and absent from that panel. \texttt{shortAndLong} is omitted for space; its results resemble \texttt{fullHorizon} (see Table~\ref{t:constr}).}
\end{figure}

Figure~\ref{fig:ufr_horizon} shows that these rankings are not specific to the thirty-day horizon. Each panel traces the average underfunding ratio under the leverage cap across holding periods of one to one hundred days, on a logarithmic scale. For \texttt{fullHorizon} and \texttt{pension}, \RI(2) delivers the lowest underfunding at every horizon, with \KRD an order of magnitude above; for \texttt{medium}, \RI(2) and \LSHD track each other closely, far below \KRD. The \texttt{longRun} panel makes the graceful degradation of the previous paragraph visible: the \RI(2) line is absent because no such portfolio exists under the cap, and \RI(1) remains close to \KRD throughout. The gaps between methods are roughly constant across horizons, so the thirty-day snapshot in Table~\ref{t:constr} is representative. In Appendix \ref{app:eval_FPY}, we repeat this simulation using the yield curve data from \citet{filipovic2025stripping} and find that the results are virtually unchanged.

\begin{table}[!htb]
\centering
\caption{Underfunding ratio under portfolio constraints.}\label{t:constr}
\begin{tabular}{lccccc}
\toprule
Liability type & \RI(0) & \RI(1) & \RI(2) & \LSHD & \KRD \\
\midrule
\multicolumn{6}{c}{\emph{Panel A: no short sales}} \\
\midrule
\texttt{fullHorizon}  & 2.43 & 0.52 & 0.46 & 0.71 & 2.27 \\
\texttt{longRun}      & 6.90 & ---  & ---  & 2.12 & 2.12 \\
\texttt{medium}       & 3.23 & 0.42 & ---  & 0.74 & 5.29 \\
\texttt{shortAndLong} & 2.10 & 0.80 & 0.75$^{\dagger}$ & 1.10 & 0.96 \\
\texttt{pension}      & 1.30 & 0.35 & 0.14 & 0.29 & 2.27 \\
\midrule
\multicolumn{6}{c}{\emph{Panel B: gross-leverage cap}} \\
\midrule
\texttt{fullHorizon}  & 2.44 & 0.96 & 0.51 & 0.71 & 2.27 \\
\texttt{longRun}      & 6.93 & 2.27 & ---  & 2.48 & 1.74 \\
\texttt{medium}       & 3.28 & 0.87 & 0.60 & 0.50 & 5.29 \\
\texttt{shortAndLong} & 2.11 & 1.09 & 0.82 & 1.06 & 0.96 \\
\texttt{pension}      & 1.30 & 0.44 & 0.15 & 0.34 & 2.27 \\
\midrule
\multicolumn{6}{c}{\emph{Panel C: collateral budget}} \\
\midrule
\texttt{fullHorizon}  & 2.44 & 1.00 & 0.55 & 0.66 & 2.27 \\
\texttt{longRun}      & 6.93 & ---  & ---  & 2.12 & 2.12 \\
\texttt{medium}       & 3.28 & 0.96 & 0.60 & 0.53 & 5.29 \\
\texttt{shortAndLong} & 2.11 & 1.11 & 0.90 & 1.09 & 0.96 \\
\texttt{pension}      & 1.30 & 0.44 & 0.15 & 0.43 & 2.27 \\
\bottomrule
\end{tabular}
\caption*{\footnotesize Note: 95th percentile of the underfunding ratio \eqref{eq:ufr} (in percent) over a thirty-day holding period and different constraints. Blank entries are infeasible. $^{\dagger}$Feasible on $53\%$ of dates.}
\end{table}

\subsection{Dynamic hedging}\label{sec:dyn}

Although the static hedging experiment in Section \ref{subsec:eval_constr} may be informative, it only addresses the performance of various immunization methods under a one-shot instantaneous change in the yield curve. In practice, the fund manager will rebalance the portfolio over time, in which case the yield curve as well as the bond maturities change. In this section we conduct a dynamic hedging experiment under the same constraints and immunization methods as in the static case, using simulated yield curve paths, and accounting for transaction costs. We simulate 5,000 yield-curve paths from the two-factor \citet{longstaff1992interest} model with GARCH volatility, estimated on the same data and anchored at randomly drawn historical curves. In Appendix \ref{app:dyn_hedge_realized}, we repeat the experiment with realized yields from \citet{liu2021reconstructing} and find that the results in favor of \RI(2) are even stronger.

Let $\set{s_n}_{n=0}^N$ be the portfolio rebalancing dates (with the normalization $s_0=0$) and assume that the liability payment dates are contained in this set. For simplicity, let $s_n=n\Delta$ with $\Delta>0$ so the dates are evenly spaced, although this is inessential. The liability pays $f_s\ge 0$ at time $s>0$. The fund manager can use $J$ zero-coupon bonds with face value 1 and maturities $\set{t_j}_{j=1}^J$ to hedge the liability. We introduce the following notation:
\begin{align*}
    x_s(t)&=\text{cumulative discount rate for term $t$ at time $s$},\\
    P_s&=\text{present value of liability at time $s$},\\
    V_s&=\text{net asset value (NAV) of fund at time $s$},\\
    z_s=(z_{sj})&=\text{immunizing portfolio at time $s$},\\
    C_s&=\text{cash position at time $s$},\\
    R_s&=\text{gross short rate at time $s$}.
\end{align*}

We now describe how to calculate these quantities recursively. At time $s$, the present value of the liability (after coupon payment) is
\begin{equation*}
    P_s\coloneqq \sum_{n:s_n>s}\e^{-x_s(s_n-s)}f_{s_n}.
\end{equation*}
Note that at time $s$, the remaining term of the $n$-th payment is $s_n-s$ and we only retain future payments in the sum. Let $s^-=s-\Delta$ denote the previous rebalancing date. The NAV of the fund consists of the present value of the bond and cash positions carried over from the previous period minus the current liability payment, which is
\begin{equation*}
    V_s\coloneqq \underbrace{R_{s^-}C_{s^-}}_\text{cash}+\underbrace{\sum_{j=1}^J z_{s^-j}\e^{-x_s(t_j-\Delta)}}_\text{bond}-\underbrace{f_s}_\text{liability}.
\end{equation*}
Here, note that the cash position earns a (predetermined) gross return $R_{s^-}$, and the zero-coupon bonds have shorter maturities $t_j-\Delta$ because time has passed. The equity (asset minus liability) is therefore
\begin{align}
    E_s&\coloneqq V_s-P_s\notag \\
    &=R_{s^-}C_{s^-}+\sum_{j=1}^Jz_{s^-j}\e^{-(x+h)(t_j-\Delta)}-f_s-\sum_{n:s_n>s}\e^{-(x+h)(s_n-s)}f_{s_n}\notag \\
    &=R_{s^-}C_{s^-}-f_s+\sum_{j=1}^Jz_{s^-j}\e^{-(x+h)(t_j-\Delta)}-\sum_{n:s_n-\Delta -s^->0}\e^{-(x+h)(s_n-\Delta-s^-)}f_{s_n}, \label{eq:equity_s}
\end{align}
where $x=x_{s^-}$ denotes the cumulative discount rate at $s^-$ and $h=x_s-x_{s^-}$ denotes the perturbation in the cumulative discount rate. As an illustration, consider the robust immunization method introduced in Section \ref{sec:main}. The fund manager's problem at time $s^-$ is to maximize the worst case equity, where the equity is defined by $E_s$ in \eqref{eq:equity_s}. Shifting $s^-$ to $s$, the time $s$ objective function is then
\begin{equation*}
    E_{s+\Delta}(z,x+h)\coloneqq R_sC_s-f_{s+\Delta}+\sum_{j=1}^Jz_{sj}\e^{-(x+h)(t_j-\Delta)}-\sum_{n:s_n-\Delta-s>0}\e^{-(x+h)(s_n-\Delta-s)}f_{s_n},
\end{equation*}
where $x=x_s$ is the current cumulative discount rate. Because $f_{s+\Delta}$ is predetermined and $C_s$ is determined by the budget constraint and hence independent of the perturbation $h$, the dynamic hedging problem reduces to the static hedging problem discussed in Section \ref{sec:main} except that \emph{all payments need to be treated as if their maturities are reduced by $\Delta$}. This modification takes into account the passage of time and hence the reduction in bond maturities by the next rebalancing date. For example, if the time to rebalancing is one quarter, a 1-year zero-coupon bond is treated as if it is a 9-month bond. The same reduction applies to every method, and the portfolio constraints of Section \ref{subsec:eval_method} are imposed at each rebalancing date.

Given the current cumulative discount rate $x_s$, it is straightforward to apply the immunization methods to bonds and liability with maturities reduced by $\Delta$. Suppose the new (time $s$) immunizing portfolio $z_s=(z_{sj})$ is chosen. Rebalancing trades incur proportional transaction costs, and the cash position is then the difference between the NAV and the portfolio value net of these costs,
\begin{equation*}
    C_s=V_s-\sum_{j=1}^J z_{sj}\e^{-x_s(t_j)}-\kappa_s,
\end{equation*}
where $\kappa_s$ denotes the transaction costs paid at time $s$. The proportional cost is half the quoted bid-ask spread, rising from half a basis point of price at the short end to three basis points at twenty years, roughly twice the on-the-run half-spreads reported by \citet[Table~3]{Fleming2003} to reflect off-the-run trading, and in line with more recent estimates \citep{Fleming2022}. Note that although we reduce the maturities by $\Delta$ to form the portfolio, we use the actual maturities to evaluate the portfolio value and define the cash position. Initializing at $V_0=P_0$ (100\% funding), we can implement dynamic hedging by repeating this procedure for $s=\Delta,2\Delta,\dotsc$. The experiment is self-financing: after the initial date the fund receives no contributions, so any hedging bias accumulates rather than being recapitalized away. We evaluate the quality of the hedge at time $s$ using the absolute return error
\begin{equation}
    \frac{1}{P_{s^-}} \abs{V_s-P_s}, \label{eq:dyn_abs_ret}
\end{equation}
and the realized funding ratio $V_s/P_s$.

We implement dynamic hedging for the \texttt{fullHorizon} and \texttt{pension} liabilities with quarterly rebalancing over a 10-year horizon, under the gross-leverage cap, which the static results show is representative of the collateral constraint as well. We compare \RI(2), \LSHD, and \KRD; exact \HD, which is infeasible under the cap, is included without constraints as a reference in the tables. 

\begin{figure}[!htb]
    \centering
    \begin{subfigure}{0.48\linewidth}
    \includegraphics[height=5.4cm]{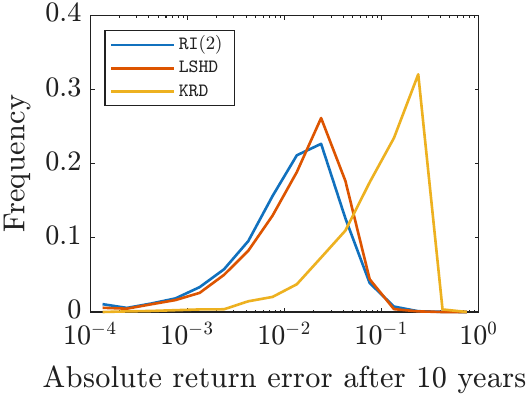}      \caption{Histogram of return error.}
    \end{subfigure}
    \begin{subfigure}{0.48\linewidth}
    \includegraphics[height=5.4cm]{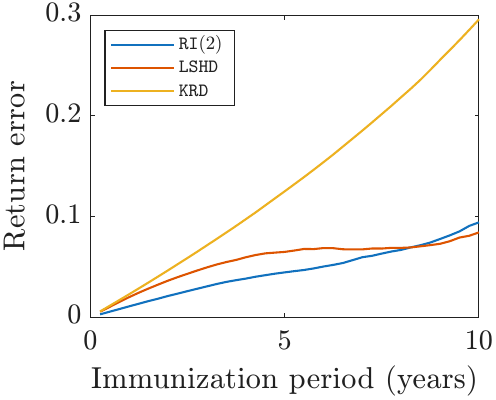}
    \caption{99th percentile of return error.}
    \end{subfigure}
    \caption{Distribution of absolute return error on simulated paths.}\label{fig:dynsim}
    \caption*{\footnotesize Note: For the \texttt{fullHorizon} liability and the simulated yield curve paths, the left panel shows the distribution of the absolute return error at the end of the 10-year immunization period and the right panel its 99th percentile throughout the period. Results for \texttt{pension} are similar.}
\end{figure}

Table~\ref{t:dyn} reports the results, and Figure~\ref{fig:dynsim} shows the distribution of hedging errors. \RI(2) attains the lowest mean error and mean squared error for both liabilities, at the lowest gross leverage and turnover among the competitive methods. \LSHD\ is the closest competitor, but it requires its entire leverage budget (gross leverage pinned at the cap of 2 for \texttt{fullHorizon}) and more than twice the turnover. \KRD fares much worse: its
mean terminal error reaches $13$--$16\%$, with a 99th percentile that grows linearly to $30\%$ over the ten years.

\KRD's poor performance reflects a measured duration gap. At every rebalancing date, \RI(2), \LSHD, and \HD\ match the liability duration essentially exactly, whereas the \KRD\ portfolio is persistently underinvested in duration: its portfolio duration averages $4.3$ years against a liability duration of $9.8$ years for \texttt{fullHorizon} (a gap of $-5.6$ years, with standard deviation $0.5$ across dates and paths), and similarly $-4.7$ years for \texttt{pension}. Because the bias is signed and \KRD\ trades very little (turnover below $0.1$ per year), the error it induces is never corrected and accumulates with the horizon, which is visible in the linear growth of its 99th percentile in Figure~\ref{fig:dynsim}.

The near-parity between \RI(2) and \LSHD\ is itself informative: the two-factor model confines every shock to the span that moment-matching methods target, making the simulation the most favorable environment for the least-squares benchmark. Indeed, for \texttt{fullHorizon} \LSHD's 99th percentile falls slightly below \RI(2)'s (Panel A of Table~\ref{t:dyn}): its positions are pinned at the leverage cap and hence bounded, which trims its extreme outcomes, at the price of twice the leverage and turnover. Transaction costs do not affect any of these rankings: they amount to at most a few basis points per year (Table~\ref{t:dyn}), and because costs are linear in the spread and \RI(2) has the lowest turnover among the competitive methods, scaling all spreads up only widens its advantage.

These rankings sharpen on realized yield curves (Appendix~\ref{app:dyn_hedge_realized}). Over 1996--2023, long-term yields declined secularly, so the signed biases that roughly cancel on trendless simulated paths instead compound in the same direction in every historical window: \RI(2) keeps the \texttt{pension} fund within $1.8\%$ of full funding over every 10-year window since 1996, at a gross leverage of one and transaction costs of a tenth of a basis point per year; \LSHD's small signed moment mismatches compound to errors roughly three times as large; and \KRD's median funding ratio declines almost linearly to $0.72$, a drift of more than a quarter of the liability value. In sum, the modest static advantage of robust immunization compounds under rebalancing: \RI(2) is the only method that is simultaneously well matched, lightly levered, and cheap to maintain.

\begin{table}[!htb]
\centering
\caption{Dynamic hedging performance.}\label{t:dyn}
\begin{tabular}{lcccccc}
\toprule
Method & Mean & p99 & MSE $\times 10^4$ & Leverage & Turnover & Cost (bp/yr) \\
\midrule
\multicolumn{7}{c}{\emph{Panel A: simulated yield curves, \texttt{fullHorizon}}} \\
\midrule
\RI(2) & 1.92 &  9.42 &   7.7 & 1.24 & 0.30 & 0.15 \\
\LSHD  & 2.13 &  8.42 &   8.1 & 2.00 & 0.69 & 0.36 \\
\KRD   & 13.0 &  29.6 & 244.6 & 1.00 & 0.03 & 0.04 \\
\HD    & 17.3 &  94.2 & 663.4 & 5.35 & 4.74 & 1.95 \\
\midrule
\multicolumn{7}{c}{\emph{Panel B: simulated yield curves, \texttt{pension}}} \\
\midrule
\RI(2) & 1.43 &  7.64 &   4.6 & 1.00 & 0.17 & 0.09 \\
\LSHD  & 1.69 &  9.06 &   6.4 & 1.84 & 0.43 & 0.22 \\
\KRD   & 15.6 &  35.4 & 355.7 & 1.00 & 0.06 & 0.03 \\
\HD    & 2.67 &  12.4 &  14.5 & 1.84 & 1.02 & 0.47 \\
\bottomrule
\end{tabular}
\caption*{\footnotesize Note: Terminal absolute return error \eqref{eq:dyn_abs_ret} after the 10-year immunization period (mean and 99th percentile in percent; MSE $\times10^4$), median gross leverage $\norm{\theta}_1$, average annual turnover (traded value over portfolio value), and average annual transaction costs in basis points. All methods are subject to the gross-leverage cap $\norm{\theta}_1\le2$ except \HD, which is infeasible under the cap and is included without constraints as a reference.}
\end{table}

\section{Conclusion}\label{sec:conclusion}

Financial institutions manage interest rate risk at the level of the balance sheet, but available bonds often cannot replicate their liability payments. We develop robust immunization for this setting. Given the liability, current yield curve, tradable bonds, and portfolio constraints, we select the feasible portfolio that performs best under the worst yield curve change in a specified stress set. We prove that this portfolio exists under weak conditions and derive a tight first-order bound on its worst equity loss. The bound separates liability value, stress size, and the exposure left after the best feasible hedge. It therefore measures the economic capital buffer needed to absorb the specified stress. Our framework accommodates general deterministic liabilities and portfolio constraints and includes duration and convexity matching as special cases.

We also provide practical ways to construct the portfolio. The general problem is a finite convex program. With an $\ell^2$ stress set and linear equality constraints, it has a closed-form constrained generalized least squares solution. A portfolio representation shows how adding stress directions can reduce the influence of unstable exact matching portfolios and thereby limit leverage. This mechanism differs from imposing leverage or collateral constraints, although those constraints can also regularize the portfolio by eliminating extreme positions.

Our evidence from U.S. Treasury yield curves shows why these features matter. In static tests based on realized yield changes, robust duration and convexity immunization delivers the best or nearly best downside performance for the four liability structures for which those matches are feasible under a leverage cap. In quarterly rebalancing experiments, it achieves the lowest mean and mean squared terminal errors for both liabilities that we study while using less leverage and turnover than the closest feasible benchmark. Its advantage is larger on historical paths because small signed mismatches compound when yields move persistently, and transaction costs do not alter the rankings. The results also reveal a limit: when the liability duration exceeds the maturity of the longest available bond, portfolio constraints can prevent the desired matches, and robust immunization need not dominate methods that concentrate on the remaining duration gap.

Our analysis focuses on small yield changes, deterministic liability cash flows, and interest rate risk. The numerical exercises examine performance under larger realized and simulated movements, but they do not establish global optimality. Extending the framework to stochastic liabilities, credit and liquidity risk, and endogenous trading costs would broaden its scope. Within its present scope, the main conclusion is simple: institutions should assess interest rate risk by the worst loss of the best feasible asset and liability portfolio, not by asset duration alone.

\section*{Acknowledgements}

The authors used large language model tools only in a supportive role to assist with coding, language editing, and manuscript organization. The authors reviewed and revised all AI-assisted material and take full responsibility for the paper's arguments, analysis, accuracy, and conclusions.

\appendix

\section{Proofs}\label{sec:proof}

\subsection{Proof of Proposition \ref{prop:minmax}}

Let us first show that $\cW$ in \eqref{eq:cW} is compact, convex, and contains 0 in the interior. Clearly $0\in \cW$. Since $w\mapsto G'w$ is linear (hence continuous) and $G'0=0$ is an interior point of $[-1,1]^N$, 0 is an interior point of $\cW$. Since $\cW$ is defined by weak linear inequalities, it is closed and convex. Let us show compactness. By Assumption \ref{asmp:H}, $H$ has full row rank, and so does $G$. Take $n_1,\dots,n_I$ such that the $I\times I$ matrix $\tilde{G}\coloneqq (g_{i,n_j})$ is invertible. Define
\begin{equation*}
    \tilde{\cW}\coloneqq \set{w\in \R^I:\tilde{G}'w\in [-1,1]^I}=(\tilde{G}')^{-1}[-1,1]^I.
\end{equation*}
Since $\tilde{\cW}$ is defined by a subset of inequalities that define $\cW$, clearly we have $\cW\subset \tilde{\cW}$. Furthermore, $\tilde{\cW}$ is compact because it is the image of the compact set $[-1,1]^I$ under the linear (hence continuous) map $(\tilde{G}')^{-1}:\R^I\to \R^I$. Therefore $\cW \subset\tilde{\cW}$ is compact.

Next, let us show that the minmax problem \eqref{eq:VIZ} has a solution $(z^*,w^*)\in \cZ\times \cW$. Since $\cW$ is nonempty and compact and $w\mapsto \seq{w,Az-b}$ is linear (hence continuous),
\begin{equation}
    M(z)\coloneqq \max_{w\in \cW}\seq{w,Az-b}
\label{eq:Mz}
\end{equation}
exists. Because $\cW$ is compact and the objective in \eqref{eq:Mz} is jointly continuous in $(z,w)$, the value function $M$ is continuous. Furthermore, since $0\in \cW$, we have $M(z)\ge 0$ and hence $V_I(\cZ)=\inf_{z\in \cZ}M(z)\ge 0$. Let $\norm{\cdot}_2$ denote the $\ell^2$ (Euclidean) norm. Since 0 is an interior point of $\cW$, there exists $\epsilon>0$ such that $w\in \cW$ whenever $\norm{w}_2\le \epsilon$. If $Az\neq b$, setting $w=\epsilon\frac{Az-b}{\norm{Az-b}_2}$, we obtain
\begin{equation}
    M(z)\ge \seq{\epsilon\frac{Az-b}{\norm{Az-b}_2},Az-b}=\epsilon\norm{Az-b}_2. \label{eq:Mzlb}
\end{equation}
Note that the lower bound \eqref{eq:Mzlb} is valid even if $Az=b$.

To bound \eqref{eq:Mzlb} from below, let us show that 
\begin{equation}
    \norm{Az-b}_2=\norm{A_+z-b_+}_2 \label{eq:l2equiv}
\end{equation}
when $z\in \cZ$. Using the definition \eqref{eq:Abplus}, it suffices to show that $a_0z-1=0$ if $z\in \cZ$. But since by Assumption \ref{asmp:Z} value matching holds, dividing \eqref{eq:value_match} by $P(x)$ and using \eqref{eq:aij} for $i=0$ (hence $h_0\equiv 1$), we obtain
\begin{equation*}
    1=\frac{1}{P(x)}\sum_{j=1}^J z_jP_j(x)=\sum_{j=1}^J a_{0j}z_j=a_0z,
\end{equation*}
which implies \eqref{eq:l2equiv}. Define $m\coloneqq \min_{\norm{z}_2=1}\norm{A_+z}_2$, which is achieved because $\norm{z}_2=1$ is a nonempty compact set and $z\mapsto \norm{A_+z}_2$ is continuous. Since by assumption $A_+$ has full column rank, we have $A_+z=0$ only if $z=0$, so $m>0$. Therefore it follows from \eqref{eq:Mzlb} and \eqref{eq:l2equiv} that for any $z\in \cZ$,
\begin{equation}
    M(z)\ge \epsilon\norm{Az-b}_2=\epsilon\norm{A_+z-b_+}_2\ge \epsilon(m\norm{z}_2-\norm{b_+}_2)\to\infty \label{eq:Mzlim}
\end{equation}
as $\norm{z}_2\to\infty$, so we may restrict the minimization of $M(z)$ to a compact subset of $\cZ$. Since $M(z)$ is continuous, the minmax value $V_I(\cZ)$ is achieved.

Finally, let us show that $z\in \cZ$ achieves $V_I(\cZ)=0$ if and only if $A_+z=b_+$. If $A_+z=b_+$, then $Az=b$ so clearly $M(z)=0$ and $V_I(\cZ)=0$. If $V_I(\cZ)=0$, then for any $z\in \cZ$ with $M(z)=V_I(\cZ)=0$, \eqref{eq:Mzlb} and \eqref{eq:l2equiv} imply $\norm{A_+z-b_+}_2=0$ and therefore $A_+z=b_+$. \hfill \qedsymbol

\subsection{Proof of Proposition \ref{prop:invariance}}

Suppose that $\spn\{\tilde{h}_i\}_{i=1}^I=\spn\set{h_i}_{i=1}^I$. Since $\set{h_i}_{i=1}^I$ span $\{\tilde{h}_i\}_{i=1}^I$, there exists an $I\times I$ matrix $C=(c_{ij})$ such that $\tilde{h}_i=\sum_{j=1}^I c_{ij}h_j$. Since $\set{h_i}_{i=1}^I$ are linearly independent, $C$ is unique. Since $\{\tilde{h}_i\}_{i=1}^I$ also span $\set{h_i}_{i=1}^I$, $C$ must be invertible. Then $\tilde{H}=CH$, $\tilde{A}=CA$, $\tilde{b}=Cb$, $\tilde{G}=CG$, so setting $w=C'\tilde{w}$, we obtain
\begin{equation*}
    \tilde{M}(z)\coloneqq \sup_{\tilde{w}:\tilde{G}'\tilde{w}\in [-1,1]^N}\seq{\tilde{w},\tilde{A}z-\tilde{b}}=\sup_{w:G'w\in [-1,1]^N}\seq{w,Az-b}\eqqcolon M(z).
\end{equation*}
Therefore the minimizers of $M$ and $\tilde{M}$ agree and the conclusion holds. \hfill \qedsymbol

\subsection{Proof of Theorem \ref{thm:RobIm}}

To prove Theorem \ref{thm:RobIm}, we recall Taylor's theorem with the integral form for the remainder term.

\begin{lem}[Taylor's theorem]\label{lem:Taylor}
Let $f\in C^{n+1}[0,1]$, so $f:[0,1]\to \R$ is $n+1$ times continuously differentiable. Then
\begin{equation}
    f(1)=\sum_{k=0}^n\frac{f^{(k)}(0)}{k!}+\int_0^1 f^{(n+1)}(s)\frac{(1-s)^n}{n!}\diff s. \label{eq:taylor}
\end{equation}
\end{lem}

\begin{proof}[Proof of Theorem \ref{thm:RobIm}]
For any $x,h\in \R$, define $f:[0,1]\to \R$ by $f(s)=\e^{-x-sh}$. Applying Lemma \ref{lem:Taylor} for $n=1$, we obtain
\begin{equation*}
    \e^{-x-h}=\e^{-x}-\e^{-x}h+\int_0^1(1-s)\e^{-x-sh}h^2\diff s.
\end{equation*}
Setting $x=x(t)$ and $h=h(t)$ for $x,h\in \cX$ and integrating both sides on $[0,T]$ with respect to $F$, we obtain
\begin{multline*}
    \int_0^T \e^{-x(t)-h(t)}\diff F(t)=\int_0^T \e^{-x(t)}\diff F(t)-\int_0^T \e^{-x(t)}h(t)\diff F(t)\\
    +\int_0^T\int_0^1(1-s)\e^{-x(t)-sh(t)}h(t)^2\diff s\diff F(t).
\end{multline*}
Using the definition of $P$ and $P'$, we obtain
\begin{equation}
    P(x+h)=P(x)+P'(x)h+\int_0^T\int_0^1(1-s)\e^{-x(t)-sh(t)}h(t)^2\diff s\diff F(t).\label{eq:taylor2}
\end{equation}
A similar equation holds for each $P_j$. Hence for any $z=(z_j)\in \R^J$ we have
\begin{equation}
    E(z,x+h)=\sum_{j=1}^J z_jP_j(x+h)-P(x+h)=E_0+E_1+E_2,\label{eq:approx_error}
\end{equation}
where
\begin{subequations}
\begin{align}
    E_0&\coloneqq \sum_{j=1}^J z_jP_j(x)-P(x),\label{eq:E0} \\
    E_1&\coloneqq \left(\sum_{j=1}^J z_jP_j'(x)-P'(x)\right)h,\label{eq:E1} \\
    E_2&\coloneqq \int_0^T\int_0^1(1-s)\e^{-x(t)-sh(t)}h(t)^2\diff s\diff \left(\sum_{j=1}^J z_jF_j(t)-F(t)\right).\label{eq:E2}
\end{align}
\end{subequations}

Since $\cZ$ satisfies value matching by Assumption \ref{asmp:Z}, we have $E_0=0$ by \eqref{eq:E0}. Inspection of Assumption \ref{asmp:H}, \eqref{eq:cH_Delta}, and \eqref{eq:cW} reveals that any $h\in \cH_I(\Delta)$ can be expressed as $h=\Delta\sum_{i=1}^I w_ih_i$ for some $w\in \cW$. Using \eqref{eq:E1}, \eqref{eq:aij}, and \eqref{eq:bi}, we obtain
\begin{equation}
    E_1=\left(\sum_{j=1}^J z_jP_j'(x)-P'(x)\right)h=-\Delta P(x)\seq{w,Az-b}.\label{eq:E1w}
\end{equation}

To bound $E_2$, note that the last integral in \eqref{eq:taylor2} is nonnegative because $1-s\ge 0$ on $s\in [0,1]$ and $F$ is increasing. Furthermore, it can be bounded above by
\begin{equation*}
    \int_0^T\int_0^1(1-s)\e^{-x(t)+\norm{h}_\infty}\norm{h}_\infty^2\diff s\diff F(t)=\frac{1}{2}\norm{h}_\infty^2\e^{\norm{h}_\infty}P(x),
\end{equation*}
where $\norm{h}_{\infty} = \max_n \abs{h(t_n)}$. (Recall that we only put restriction on $h$ at the payout dates.) Therefore $E_2$ in \eqref{eq:E2} can be bounded as
\begin{equation}
    \frac{1}{2}\norm{h}_\infty^2\e^{\norm{h}_\infty}\left(\sum_{z_j<0}z_jP_j(x)-P(x)\right) \le E_2\le \frac{1}{2}\norm{h}_\infty^2\e^{\norm{h}_\infty}\sum_{z_j\ge 0}z_jP_j(x).\label{eq:E2bd}
\end{equation}
Using \eqref{eq:value_match} and \eqref{eq:theta}, we obtain
\begin{align}
    P(x)-\sum_{z_j<0}z_jP_j(x)&=\sum_{z_j\ge 0}z_jP_j(x)=\frac{1}{2}\left(P(x)+\sum_{j=1}^J\abs{z_j}P_j(x)\right)\notag\\
    &=\frac{1}{2}P(x)\left(1+\sum_{j=1}^J\abs{\theta_j}\right)=\frac{1}{2}P(x)(1+\norm{\theta}_1).\label{eq:Pjsum}
\end{align}
Noting that $\norm{h}_\infty\le \Delta T$ for $h\in \cH_I(\Delta)$, it follows from \eqref{eq:E2bd} and \eqref{eq:Pjsum} that\begin{equation}
    \abs{E_2}\le \frac{1}{4}\Delta^2T^2\e^{\Delta T}P(x)(1+\norm{\theta}_1).\label{eq:E2w}
\end{equation}
Combining \eqref{eq:approx_error}, $E_0=0$, \eqref{eq:E1w}, and \eqref{eq:E2w}, we obtain
\begin{multline}
    -\seq{w,Az-b}-\frac{1}{4}\Delta T^2\e^{\Delta T}(1+\norm{\theta}_1)\\
    \le \frac{1}{\Delta P(x)}E(z,x+h)
    \le -\seq{w,Az-b}+\frac{1}{4}\Delta T^2\e^{\Delta T}(1+\norm{\theta}_1).\label{eq:loss_bound1}
\end{multline}
Since by \eqref{eq:theta} $\theta_j$ is proportional to $z_j$, there exists some constant $c(x)>0$ that depends only on $x$ such that $\norm{\theta}_1\le c(x)\norm{z}_2$. Therefore minimizing \eqref{eq:loss_bound1} over $w\in \cW$, it follows from the definition of $M(z)$ in \eqref{eq:Mz} that
\begin{multline}
    -M(z)-\frac{1}{4}\Delta T^2\e^{\Delta T}(1+c(x)\norm{z}_2)\\
    \le \frac{1}{\Delta P(x)}\inf_{h\in \cH_I(\Delta)}E(z,x+h)
    \le -M(z)+\frac{1}{4}\Delta T^2\e^{\Delta T}(1+c(x)\norm{z}_2).\label{eq:loss_bound2}
\end{multline}
Let $m,\epsilon>0$ be as in the proof of Proposition \ref{prop:minmax} and take $\bar{\Delta}>0$ such that $\epsilon m=\frac{1}{4}\bar{\Delta} T^2\e^{\bar{\Delta} T}c(x)$. Then if $0<\Delta<\bar{\Delta}$, by \eqref{eq:Mzlim} both sides of \eqref{eq:loss_bound2} tend to $-\infty$ as $\norm{z}_2\to\infty$. Therefore when we take the supremum of \eqref{eq:loss_bound2} with respect to $z\in \cZ$, we may restrict it to some compact subset $\cZ'\subset \cZ$. Therefore there exists a constant $c'>0$ such that
\begin{equation*}
    -M(z)-c'\Delta\le \frac{1}{\Delta P(x)}\inf_{h\in \cH_I(\Delta)}E(z,x+h)\le -M(z)+c'\Delta
\end{equation*}
for all $z\in \cZ'$ and $\Delta\in (0,\bar{\Delta})$. Taking the supremum over $z\in \cZ$ (which is achieved in $\cZ'$) and letting $\Delta\to 0$, by the definition of $V_I(\cZ)$ in \eqref{eq:VIZ}, we obtain \eqref{eq:maxmin0}.

To show the error estimate \eqref{eq:RobIm_error_bd}, let $z^*\in \cZ$ be a solution to the minmax problem \eqref{eq:VIZ}. It follows from \eqref{eq:loss_bound1} that
\begin{equation*}
    \frac{1}{\Delta P(x)}\abs{E(z^*,x+h)}
    \le \abs{\seq{w,Az^*-b}}+\frac{1}{4}\Delta T^2\e^{\Delta T}(1+\norm{\theta}_1).
\end{equation*}
Taking the supremum over $w\in \cW$ and noting that $\cW$ is symmetric ($w\in \cW$ implies $-w\in \cW$), it follows from the definition of $V_I(\cZ)$ in \eqref{eq:VIZ} that \eqref{eq:RobIm_error_bd} holds.
\end{proof}

\subsection{Proof of Proposition \ref{prop:maxmin_global}}

Suppose that the liability has maturity $s$ with face value 1. Then the value of the liability is
\begin{equation*}
    P(x)=\int_0^T \e^{-x(t)}\diff F(t)=\e^{-x(s)}.
\end{equation*}
Let $z^*=A_+^{-1}b_+$ be the immunizing portfolio and assume $z^*\ge 0$. Take any perturbation $h\in \spn\set{h_i}_{i=1}^I$ and write $h=\sum_{i=1}^I w_ih_i$. Then the funding ratio is
\begin{equation*}
    \phi(w)\coloneqq \frac{\sum_{j=1}^J z_j^*P_j(x+h)}{P(x+h)}=\sum_{j=1}^J z_j^*\int_0^T \e^{-x(t)+x(s)-h(t)+h(s)}\diff F_j(t).
\end{equation*}
Since $z^*\ge 0$ and the exponential function is convex, $\phi(w)$ is convex in $w\in \R^I$.

Let us show that $\nabla \phi(0)=0$. To this end we compute  
\begin{align}
    \frac{\partial \phi}{\partial w_i}(0)&=\sum_{j=1}^J z_j^*\int_0^T\e^{-x(t)+x(s)}(-h_i(t)+h_i(s))\diff F_j(t) \notag \\
    &=\e^{x(s)}\sum_{j=1}^J z_j^*\left(-\int_0^T \e^{-x(t)}h_i(t)\diff F_j(t) + h_i(s)\int_0^T \e^{-x(t)}\diff F_j(t)\right) \notag \\
    &=\e^{x(s)}\left(-P(x)\sum_{j=1}^J a_{ij}z_j^*+h_i(s)\sum_{j=1}^J z_j^*P_j(x)\right), \label{eq:dphi1}
\end{align}
where the last line uses \eqref{eq:aij} and \eqref{eq:Px} for each bond $j$. Using value matching \eqref{eq:value_match} and the fact that the liability is a zero-coupon bond, we obtain
\begin{equation}
    h_i(s)\sum_{j=1}^J z_j^*P_j(x)=h_i(s)P(x)=\e^{-x(s)}h_i(s)=\int_0^T \e^{-x(t)}h_i(t)\diff F(t)=P(x)b_i, \label{eq:dphi2}
\end{equation}
where the last equality uses \eqref{eq:bi}. Combining \eqref{eq:dphi1} and \eqref{eq:dphi2}, we obtain
\begin{equation}
    \nabla \phi(0)=b-Az^*=0. \label{eq:dphi}
\end{equation}
Since $\phi$ is convex, it follows that $\phi(w)\ge \phi(0)=1$ for all $w$, which implies $E(z^*,x+h)\ge 0$. \hfill \qedsymbol

\printbibliography

\newpage

\begin{center}
    {\Large \textbf{Online Appendix (Not for publication)}}
\end{center}

\section{Additional proofs}

\subsection{Proof of Proposition \ref{prop:RobIml2}}

To solve the inner maximization of \eqref{eq:VIZp}, fix $z$ and let $c=Az-b$. Then the optimization problem reduces to
\begin{align*}
    &\maximize && \seq{c,w}\\
    &\st && \frac{1}{2}w'GG'w\le \frac{1}{2}.
\end{align*}
If $c=0$, the maximum value is 0, and any $w$ is optimal. Therefore assume $c\neq 0$ and define the Lagrangian by
\begin{equation*}
    L(w,\lambda)\coloneqq \seq{c,w}+\frac{\lambda}{2}(1-w'GG'w),
\end{equation*}
where $\lambda\ge 0$ is the Lagrange multiplier. Since by Assumption \ref{asmp:H} the matrix $G$ has full row rank, $GG'$ is positive definite. Then the Slater constraint qualification is satisfied and we can apply the Karush-Kuhn-Tucker theorem. The first-order condition is
\begin{equation*}
    0=\nabla_w L=c-\lambda GG'w\iff w=\frac{1}{\lambda}(GG')^{-1}c.
\end{equation*}
The complementary slackness condition is
\begin{equation*}
    1=w'GG'w=\frac{1}{\lambda^2}c'(GG')^{-1}c\iff \lambda=\sqrt{c'(GG')^{-1}c}.
\end{equation*}
Therefore the maximum value is
\begin{equation*}
    \seq{c,w}=\frac{c'(GG')^{-1}c}{\lambda}=\sqrt{c'(GG')^{-1}c},
\end{equation*}
which is also valid when $c=0$.

Noting that $c=Az-b$, the outer minimization reduces to
\begin{align*}
    &\minimize && \frac{1}{2}\seq{Az-b,(GG')^{-1}(Az-b)}\\
    &\st && Rz=r.
\end{align*}
Let $\mu\in \R^M$ be the Lagrange multiplier for the equality constraint and define the Lagrangian by
\begin{equation*}
    L(z,\mu)\coloneqq \frac{1}{2}\seq{Az-b,(GG')^{-1}(Az-b)}+\mu'(r-Rz).
\end{equation*}
Letting $\tilde{A}\coloneqq (GG')^{-1} A$, the first-order condition is
\begin{equation*}
    \tilde{A}'(Az-b)-R'\mu=0\iff z=(\tilde{A}'A)^{-1}(\tilde{A}'b+R'\mu),
\end{equation*}
where we used the fact that $\tilde{A}'A=A'(GG')^{-1}A$ is invertible because $A$ has full column rank. The equality constraint implies
\begin{equation*}
    r=Rz=R(\tilde{A}'A)^{-1}(\tilde{A}'b+R'\mu)\iff \mu=[R(\tilde{A}'A)^{-1}R']^{-1}(r-R(\tilde{A}'A)^{-1}\tilde{A}'b),
\end{equation*}
where we used the fact that $R(\tilde{A}'A)^{-1}R'$ is invertible because $R$ has full row rank. Therefore the solution to the minmax problem \eqref{eq:VIZp} for $p=2$ is
\begin{equation*}
    z=(\tilde{A}'A)^{-1}\tilde{A}'b+(\tilde{A}'A)^{-1}R'[R(\tilde{A}'A)^{-1}R']^{-1}(r-R(\tilde{A}'A)^{-1}\tilde{A}'b). \eqno \qedsymbol
\end{equation*}

\subsection{Proof of Proposition \ref{prop:cgls}}\label{app:proof_cgls}
We prove Proposition \ref{prop:cgls} by deriving a general result that expresses a constrained GLS estimator in terms of elemental estimates. We first derive the analogous result for the constrained OLS estimator, which might be of independent interest. In this section we always assume $A \in \R^{I \times J}$, $b \in \R^{I}$ with $I \ge J$, and $\rank(A) = J$. Furthermore, $[M]$ denotes the set of integers $\set{1,\dots,M}$, and $s$ denotes any subset of size $J$ from the set $\set{1,\dots,I}$. 

\begin{prop}\label{prop:jacobi}
Let $z_{cls} \in \mathbb{R}^J$ be the solution to the constrained least squares problem $\min_z \norm{b - Az}_2^2$ subject to the linear constraints $Rz = r$, where $R \in \mathbb{R}^{M \times J}$ has full row rank and $r \in \R^M$. Define the augmented matrices 
\begin{equation*}
A_+ = \begin{bmatrix}
 R\\
 A
\end{bmatrix} \in \R^{(M + I) \times J} \qquad
b_+ = \begin{bmatrix}
r \\
b
\end{bmatrix} \in \R^{M + I},
\end{equation*}
and let $z(s) = A_+(s)^{-1} b_+(s) \in \R^J$ denote the elemental estimate based on the rows of $A_+$ and $b_+$ that are in $s$. Then, 
\begin{align}\label{eq:cls}
z_{\text{cls}} &= \E(z(S)\mid [M] \subset S)  = \sum_{[M] \subset s}  \frac{\det[A_+(s)]^2 }{\sum_{[M] \subset s} \det[A_+(s)]^2}  z(s)  \\
&\eqqcolon \sum_{[M] \subset s} \lambda(s) z(s), \nonumber
\end{align}
where 
\begin{equation*}
\lambda(s) = \frac{\det[A_+(s)]^2 }{\sum_{[M] \subset s} \det[A_+(s)]^2} \ge 0 \quad \text{and } \sum_{[M] \subset s} \lambda(s) = 1.
\end{equation*}
\end{prop}

In the proof we use the generalized Cauchy-Binet formula \citep[Appendix B]{chapman2018classical}.  
\begin{lem}[Generalized Cauchy-Binet formula]\label{lem:cauchy}
Let $C \in \mathbb{R}^{m \times n}$ and $D \in \mathbb{R}^{n \times m}$, and let $T$ denote the set $\set{n-j+1,\dots,n}$. Let $s$ denote any subset of $\set{1,\dots,n}$ of cardinality $m$. Denote $C(s) \in \mathbb{R}^{m \times \abs{s}}$ by the matrix $C$ with column index in $s$ and $D(s) \in \mathbb{R}^{\abs{s} \times m}$ by the matrix $D$ with row index in $s$. Then,  
\begin{equation*}
\sum_{T \subset s} \det\lr{C(s)} \det\lr{D(s)} = (-1)^j \det\begin{bmatrix}
0_{j\times j} & D(s) \\
C(s) & C([n-j]) D([n-j])
\end{bmatrix}.
\end{equation*}
\end{lem}

\begin{proof}[Proof of Proposition \ref{prop:jacobi}]
By construction the expression for $z_{\text{cls}}$ in \eqref{eq:cls} satisfies $R z_{\text{cls}} = r$ since the sum is taken only over rows that also contain the constraints. Using Lemma \ref{lem:cauchy} we can write
\begin{align*}
&\sum_{[M] \subset s} \det[A_+(s)]^2 = \sum_{[M] \subset s} \det[A_+(s)'] \det[A_+(s)]\\
&= (-1)^M \det \begin{bmatrix}
A_+([M])' & A_+\lro{\set{M+1,\dots,I}}' A_+\lro{\set{M+1,\dots,I}}\\
0_{M \times M} & A_+\lro{[M]}
\end{bmatrix} \\
&= (-1)^M \det \begin{bmatrix}
R' &  A'A\\
0_{M\times M} & R
\end{bmatrix}\\
&= (-1)^M \det\lr{A'A} \det\lr{R (A'A)^{-1} R'}.
\end{align*}
The final equality follows from the determinant formula for block matrices. Now we define $z^{j}(s)$ to be the $j$th element of $z(s)$. Using Cramer's rule we have $z^{j}(s) = \det\lr{A_+^j(s)}/\det\lr{A_+(s)}$, where $A_+^j(s)$ is $A_+(s)$ except that the $j$th column is replaced by $b_+(s)$. Similarly $R^j$ is defined such that the $j$th column of $R$ is replaced by $r$. It then follows that  
\begin{align*}
&\sum_{[M] \subset s}  \det[A_+(s)]^2 z^j(s) = \sum_{[M] \subset s}  \det\lr{A_+(s)'} \det\lr{A_+^j(s)}\\
&= (-1)^M \det \begin{bmatrix}
A_+([M])' &  A_+\lro{\set{M+1,\dots,I}}' A_+^j\lro{\set{M+1,\dots,I}}\\
0_{M \times M} & A_+^j\lro{[M]}
\end{bmatrix}
\\
&= (-1)^M \det \begin{bmatrix}
R' & A' A^j\\
0_{M\times M} & R^j
\end{bmatrix}\\
&= (-1)^M \det\lr{A' A^j} \det\lr{R^j (A'A^j)^{-1} R'}.
\end{align*}
In conclusion we have shown that the $j$th element of $z_{\text{cls}}$ in \eqref{eq:cls} can be expressed as
\begin{equation}
    z_{\text{cls}}^j = \frac{\det\lr{A' A^j} \det\lr{R^j (A'A^j)^{-1} R'}}{\det\lr{A'A} \det\lr{R (A'A)^{-1} R'}}= z_{\text{ols}}^j \frac{\det\lr{R^j (A'A^j)^{-1} R'}}{\det\lr{R (A'A)^{-1} R'}}, \label{eq:RAA}
\end{equation}
where $z_{\text{ols}}^j$ is the $j$th element of the unconstrained OLS solution $\min_z \norm{Az - b}_2^2$. The latter observation follows immediately from Cramer's rule applied to the first-order condition $A'A z_{\text{ols}} = A' b$. 

It remains to verify that the solution in \eqref{eq:RAA} concurs with the constrained least squares solution at the top of Proposition \ref{prop:jacobi}. Let $\mu \in \R^M$ be the Lagrange multiplier for the equality constraint and consider the associated Lagrangian 
\begin{equation*}
L(z,\mu) = \frac{1}{2}\norm{Az - b}_2^2 + \mu' \lro{Rz - r}.
\end{equation*}
The first order conditions for this problem imply 
\begin{equation*}
\begin{bmatrix}
A'A & R'\\
R & 0
\end{bmatrix}
\begin{bmatrix}
    z_{\text{cls}}\\
    \mu
\end{bmatrix} = 
\begin{bmatrix}
A' b\\
r
\end{bmatrix}.
\end{equation*}
Using Cramer's rule to get the $j$th element of $z_{\text{cls}}$ we get
\begin{equation*}
z_{\text{cls}}^j = \frac{\det \begin{bmatrix}
A' A^j & R'\\
R^j & 0
\end{bmatrix} }{\det\begin{bmatrix}
A'A & R'\\
R & 0
\end{bmatrix}}
= \frac{\det\lr{A' A^j} \det\lr{R^j (A'A^j)^{-1} R'}}{\det\lr{A'A} \det\lr{R (A'A)^{-1} R'}}. \qedhere
\end{equation*}
\end{proof}

\begin{prop}\label{prop:cgls_general}
Consider the constrained GLS solution 
\begin{equation}\label{eq:cgls}
z_{cgls} \coloneqq \argmin_z \seq{Az - b,\Omega^{-1}(Az - b)} \quad \text{s.t. } Rz = r.
\end{equation}
Define $\tilde{A} \coloneqq \Omega^{-1} A$ and let $\tilde{A}_+' = [R, \tilde{A}]'$. Let $A_+$ and $z(s)$ be the same as in Proposition \ref{prop:jacobi}. Then the solution in \eqref{eq:cgls} can be expressed as
\begin{equation*}
z_{cgls} = \frac{\sum_{[M] \subset s} \det\lr{\tilde{A}_+(s)} \det\lr{A_+(s) } z(s) }{\sum_{[M] \subset s} \det\lr{\tilde{A}_+(s)} \det\lr{A_+(s) }} \eqqcolon \sum_{[M] \subset s} \lambda(s) z(s),
\end{equation*}
where 
\begin{equation*}
\lambda(s) = \frac{\det\lr{\tilde{A}_+(s)} \det\lr{A_+(s) }}{\sum_{[M] \subset s} \det\lr{\tilde{A}_+(s)} \det\lr{A_+(s) }} \quad  \text{with} \quad \sum_{[M] \subset s} \lambda(s) = 1.
\end{equation*}
\end{prop}
\begin{rem}
In this case $\lambda(\cdot)$ does not necessarily define a conditional probability measure since it can take on negative values.
\end{rem}

\begin{proof}
The steps are similar to the proof of Proposition \ref{prop:jacobi}. The first order conditions for the optimization problem in \eqref{eq:cgls} imply
\begin{equation*}
\begin{bmatrix}
A' \Omega^{-1} A & R'\\
R & 0
\end{bmatrix}
\begin{bmatrix}
z_{cgls}\\
\mu
\end{bmatrix}
= \begin{bmatrix}
A' \Omega^{-1} b\\
r
\end{bmatrix}.
\end{equation*}
Applying Cramer's rule yields
\begin{align*}
z_{cgls}^j &= \frac{\det \begin{bmatrix}
A' \Omega^{-1} A^j & R'\\
R^j & 0
\end{bmatrix}}{\det \begin{bmatrix}
A' \Omega^{-1} A & R'\\
R & 0
\end{bmatrix}}\\
&= \frac{\det\lr{A' \Omega^{-1} A^j} \det \lr{R^j (A' \Omega^{-1} A^j)^{-1} R'}   }{\det\lr{A' \Omega^{-1} A} \det \lr{R (A' \Omega^{-1} A)^{-1} R'}  }\\
&= \frac{\det\lr{\tilde{A}' A} \det \lr{R^j (\tilde{A}' A^j)^{-1} R'}   }{\det\lr{\tilde{A}' A} \det \lr{R (\tilde{A}' A)^{-1} R'}  }.
\end{align*}
The rest of the proof is identical to Proposition \ref{prop:jacobi}.
\end{proof}
Finally we prove Proposition \ref{prop:cgls} as a special case of Proposition \ref{prop:cgls_general}.
\begin{proof}[Proof of Proposition \ref{prop:cgls}]
Due to the basis invariance of the robust immunization portfolio in \eqref{eq:z_l2}, we can use the polynomial basis $h_i(t) = t^i$. Therefore, the first $J-1$ rows of $A$ are equal to $A_{\texttt{HD}}$. The result now follows from Proposition \ref{prop:cgls_general} with $\Omega = GG'$, $R = a_0$, and $r = 1$.
\end{proof}

\subsection{Proof of Theorem \ref{thm:RobIm1}}

Because the proof is similar to that of Theorem \ref{thm:RobIm}, we only provide a sketch.

By assumption, $\cZ_1$ in \eqref{eq:cZ1} is nonempty, and it is clearly closed. Hence by Proposition \ref{prop:minmax} the minmax value $V_I^p(\cZ_1)$ defined by \eqref{eq:VIZp} is achieved by some $z^*\in \cZ_1$. Inspection of Assumption \ref{asmp:H}, \eqref{eq:cH_Delta1}, and \eqref{eq:cW} reveals that any $h\in \cH_I^p(\Delta_1,\Delta_2)$ can be expressed as $h=\Delta_1vh_1+\Delta_2\sum_{i=1}^I w_ih_i$ for some $w\in \cW^p$ and $v\in \R$ with $\abs{v}\le  1/\norm{h_1(t)/t}_{p,t} \eqqcolon \bar{v}\in (0,\infty)$. Applying a similar argument to the derivation of \eqref{eq:loss_bound1}, we obtain
\begin{equation*}
    \frac{1}{P(x)}E(z,x+h)=-\Delta_1v(Az-b)_1-\Delta_2 \seq{w,Az-b}+O(\Delta_1^2+\Delta_2^2),
\end{equation*}
where $(Az-b)_1$ denotes the first entry of the vector $Az-b$. Minimizing both sides over $h\in \cH_I^p(\Delta_1,\Delta_2)$, we obtain
\begin{equation*}
    \inf_{h\in \cH_I^p(\Delta_1,\Delta_2)}\frac{1}{P(x)}E(z,x+h)=-\Delta_1\bar{v}\abs{(Az-b)_1}-\Delta_2 M^p(z)+O(\Delta_1^2+\Delta_2^2),
\end{equation*}
where $M^p(z) = \max_{w\in \cW^p}\seq{w,Az-b}$. Dividing both sides by $\Delta_2>0$ and letting $\Delta_2\to 0$, $\Delta_1/\Delta_2\to \infty$, and $\Delta_1^2/\Delta_2\to 0$, the objective function remains finite only if $(Az-b)_1=0$, which is equivalent to $z\in \cZ_1$. Under this condition, we have
\begin{equation*}
    \frac{1}{\Delta_2}\inf_{h\in \cH_I^p(\Delta_1,\Delta_2)}\frac{1}{P(x)}E(z,x+h)=-M^p(z)+O(\Delta_2+\Delta_1^2/\Delta_2).
\end{equation*}
Maximizing over $z\in \cZ_1$ and taking limits, we obtain \eqref{eq:maxmin1}. The proof of \eqref{eq:RobIm_error_bd1} is similar. \hfill \qedsymbol

\subsection{Proof of Proposition \ref{prop:monotoneVIZ}}

For each $I$, let $M_I(z)=\sup_{w\in \cW_I^p}\seq{w,A_Iz-b_I}$, where $A_I,b_I$ denote the sensitivity matrix and vector $A,b$ defined by \eqref{eq:aij}, \eqref{eq:bi} and $\cW_I^p$ denotes the set $\cW^p$ defined by \eqref{eq:cWp}. Suppose $I<I'$. Letting $0_N$ denote the zero vector of $\R^N$, we have $\cW_I^p\times \set{0_{I'-I}}\subset \cW_{I'}^p$, so
\begin{align*}
    M_I(z)&=\sup_{w\in \cW_I^p}\seq{w,A_Iz-b_I}=\sup_{w\in \cW_I^p\times\set{0_{I'-I}}}\seq{w,A_{I'}z-b_{I'}}\\
    &\le \sup_{w\in \cW_{I'}^p}\seq{w,A_{I'}z-b_{I'}}=M_{I'}(z).
\end{align*}
Taking the infimum over $z\in \cZ$, we obtain $V_I^p(\cZ)\le V_{I'}^p(\cZ)$. Similarly,
\begin{equation*}
    V_I^p(\cZ)=\inf_{z\in \cZ}M_I(z)\ge \inf_{z\in \cZ'}M_I(z)=V_I^p(\cZ'). \eqno\qedsymbol
\end{equation*}

\section{Two-factor term structure model}\label{sec:LS}

We use the two-factor term structure model of \citet{longstaff1992interest} to simulate yield curves.  This appendix summarizes the model and presents parameter estimates based on our yield curve data.

\subsection{Model estimation}
In \citet{longstaff1992interest}, the change in yields from time $s$ to $s'$ is given by
\begin{equation}\label{eq:LS_yields}
y_{s'}(t) - y_s(t) = \lr{-C(t) (r_{s'} - r_s) - D(t) (V_{s'} - V_s) } /t,
\end{equation}
where $r_s$ denotes the short rate at time $s$, $V_s$ denotes the short rate variance at time $s$, and $C(t), D(t)$ are time-invariant factors that only depend on the maturity $t$ and model parameters. Because the conditional variance is a latent state variable, we follow \citet{longstaff1992interest} and estimate it using the following GARCH specification:
\begin{align*}
r_{s+1} - r_s &= \alpha_0 + \alpha_1 r_s + \alpha_2 V_s + \varepsilon_{s+1},\\
\varepsilon_{s+1} &\sim \mathsf{N}(0,V_s), \\
V_s &= \beta_0 + \beta_1 r_s + \beta_2 V_{s-1} + \beta_3 \varepsilon_s^2.
\end{align*}
We estimate the parameter vector $[\alpha_0,\alpha_1,\alpha_2,\beta_0,\beta_1,\beta_2,\beta_3]'$ by maximum likelihood using monthly changes in the 3-month yield, which we take as the short rate.

Given these estimated conditional variances, the functions $C(t)$ and $D(t)$ can be estimated by GMM and expressed in terms of the model parameters 
$[\alpha,\beta,\delta,\nu]'$ as follows:
\begin{align*}
A(t) &= \frac{2 \phi}{(\delta + \phi) (\exp(\phi t) - 1) + 2 \phi},\\
B(t) &= \frac{2 \psi}{(\nu + \psi)(\exp(\psi t) - 1) + 2 \psi}, \\
C(t) &= \frac{\alpha \phi(\exp(\psi t) - 1)B(t) - \beta \psi (\exp(\phi t) - 1) A(t)}{\phi \psi (\beta - \alpha)},\\
D(t) &= \frac{\psi (\exp(\phi t) - 1)A(t) - \phi (\exp(\psi t) - 1) B(t)}{\phi \psi (\beta - \alpha)},
\end{align*}
where $\phi = \sqrt{2 \alpha + \delta^2}$ and $\psi = \sqrt{2 \beta + \nu^2}$. Table~\ref{tab:LS_estimate} presents the estimated model parameters. 
To simulate yields, we randomly select an initial yield curve from the daily yield curve data of \citet{liu2021reconstructing}.  Starting from this curve, we simulate monthly yield curve changes using \eqref{eq:LS_yields}.  Finally, we retain only the quarterly yield curves, as our application involves quarterly rebalancing.

\begin{table}[!htb]
\centering
\caption{Estimated parameters of term structure model.}
\label{tab:LS_estimate}
\begin{tabular}{l r}
\toprule
\midrule
Parameter & Estimate \\
\midrule
$\alpha_0 \times 100$ & 0.003 \\
$\alpha_1$ & 0.003 \\
$\alpha_2$ & $-0.002$ \\
$\beta_0 \times 100$  & 0.000 \\
$\beta_1 \times 100$  & 0.004 \\
$\beta_2$  & 0.407 \\
$\beta_3$  & 0.486 \\
\midrule
$\alpha$   & $-0.038$ \\
$\beta$    & 2.940 \\
$\delta$   & 1.546 \\
$\nu$      & 14.458 \\
\bottomrule
\end{tabular}
\end{table}

\section{Miscellaneous results}

\subsection{Space of cumulative discount rates}\label{app:space}

\begin{lem}\label{lem:cX_Banach}
Let $\cX=\set{x\in C[0,T]:x(0)=0}$ be the vector space of continuous functions on $[0,T]$ with $x(0)=0$. For $x\in \cX$, define $\norm{x}_\cX=\sup_{t\in [0,T]}\abs{x(t)}$. Then $(\cX,\norm{\cdot}_\cX)$ is a Banach space.
\end{lem}

\begin{proof}
It is well known that the space $C[0,T]$ endowed with the supremum norm is a Banach space (\eg~\citet[Chapter 5]{Folland1999}). Let $L_0 x = x(0)$ be the evaluation functional at 0. Since $L_0$ is a continuous map and $\set{0}$ is closed, we get that $L_0^{-1} \set{0} = \cX$ is a closed set. Since a closed subspace of a Banach space is a Banach space, the proof is complete.
\end{proof}

\begin{lem}[Polynomial basis]\label{lem:poly}
Suppose Assumption \ref{asmp:payouts} holds and $h_i$ is a polynomial of degree $i$ with $h_i(0)=0$. Then Assumption \ref{asmp:H} holds.
\end{lem}

\begin{proof}
Since $h_i$ is a polynomial of degree $i$ with $h_i(0)=0$, without loss of generality we may assume $h_i(t)=t^i$. To show that $(h_i)$ is dense in $\cX$, let $x\in \cX$, use the Stone-Weierstrass theorem \citep[p.~139]{Folland1999} to approximate $x$ by a polynomial $p$, and replace $p$ by $p-p(0)$. This polynomial vanishes at 0 and lies in $\spn (h_i)$, so $(h_i)$ is dense. By Assumption \ref{asmp:payouts}, we can choose $I$ distinct points $\set{t_{n_j}}_{j=1}^I$. Consider the $I\times I$ submatrix of $H$ defined by $\tilde{H}=(h_i(t_{n_j}))=(t_{n_j}^i)$. Dividing the $j$-th column by $t_{n_j}>0$, $\tilde{H}$ reduces to a Vandermonde matrix, which is invertible. Therefore $H$ has full row rank. The same argument applies to $G$.
\end{proof}

\begin{lem}\label{lem:bounded_linear_operator}
The Fr\'echet derivative $P'(x)$ defined by $h\mapsto \delta P(x;h)$ in \eqref{eq:gateaux} is a bounded linear operator.
\end{lem}

\begin{proof}
Clearly $P'(x)$ is a linear operator. If $h\in \cX$, then 
\begin{equation*}
    \abs{P'(x)h} \le \int_0^T \e^{-x(t)}\abs{h(t)}\diff F(t)\le \norm{h}_\cX \int_0^T \e^{-x(t)}\diff F(t),
\end{equation*}
so $P'(x)$ is a bounded linear operator with operator norm less than or equal to $\int_0^T \e^{-x(t)}\diff F(t)$.
\end{proof}

\subsection{Convexifying the set of priors}\label{app:convex}
The maxmin problem in Theorem \ref{thm:RobIm} can be interpreted as the choice of a decision maker with ambiguity-averse preferences whose set of priors consists of point masses on shocks in the stress set,
\begin{equation*}
\Pi_0(\Delta)\coloneqq \set{\delta_h: h\in \cH_I(\Delta)},
\end{equation*}
where $\delta_h$ denotes the Dirac measure putting probability one on the shock $h$. This set of priors may be considered overly dogmatic and, from a technical standpoint, is not convex: for distinct $g,h\in \cH_I(\Delta)$ and $\alpha\in (0,1)$, we have $\alpha \delta_g+(1-\alpha)\delta_h\notin \Pi_0(\Delta)$. In contrast, the axiomatization of \citet{GilboaSchmeidler1989} assumes a closed and convex set of priors. We therefore extend the set of priors to all probability measures supported on the stress set,
\begin{equation*}
\Pi(\Delta)\coloneqq \set{\pi: \text{$\pi$ is a Borel probability measure on $\cH_I(\Delta)$}},
\end{equation*}
which is convex, contains $\Pi_0(\Delta)$, and in particular contains every finite mixture $\sum_{k=1}^K \alpha_k\delta_{h^{(k)}}$ of point masses on admissible shocks $h^{(k)}\in \cH_I(\Delta)$. The following proposition shows that this extension leaves the maxmin problem unchanged.

\begin{prop}\label{prop:ambiguity}
Let everything be as in Theorem \ref{thm:RobIm}. For every $z\in \cZ$ and $\Delta>0$, we have
\begin{equation}
\inf_{h\in \cH_I(\Delta)} E(z,x+h)=\inf_{\pi\in \Pi(\Delta)} \Expect_\pi E(z,x+h), \label{eq:ambiguity}
\end{equation}
where $\Expect_\pi$ denotes the expectation with respect to $h\sim \pi$. Consequently,
\begin{equation*}
\sup_{z\in \cZ}\inf_{h\in \cH_I(\Delta)} E(z,x+h)=\sup_{z\in \cZ}\inf_{\pi\in \Pi(\Delta)} \Expect_\pi E(z,x+h).
\end{equation*}
\end{prop}

\begin{proof}
Fix $z\in \cZ$ and $\Delta>0$ and let $\underline{E}\coloneqq \inf_{h\in \cH_I(\Delta)} E(z,x+h)$, which is finite because $\norm{h}_\infty\le \Delta T$ for $h\in \cH_I(\Delta)$. Take any $\pi\in \Pi(\Delta)$. Since $E(z,x+h)\ge \underline{E}$ for all $h\in \cH_I(\Delta)$ and $\pi$ is supported on $\cH_I(\Delta)$, monotonicity of the expectation yields $\Expect_\pi E(z,x+h)\ge \underline{E}$. Taking the infimum over $\pi\in \Pi(\Delta)$ gives $\inf_{\pi\in \Pi(\Delta)} \Expect_\pi E(z,x+h)\ge \underline{E}$. Conversely, since $\Pi(\Delta)\supset \Pi_0(\Delta)$ and $\Expect_{\delta_h} E(z,x+h)=E(z,x+h)$, we obtain $\inf_{\pi\in \Pi(\Delta)} \Expect_\pi E(z,x+h)\le \underline{E}$. Therefore \eqref{eq:ambiguity} holds. The second claim follows by taking the supremum over $z\in \cZ$ on both sides of \eqref{eq:ambiguity}.
\end{proof}

Note that, unlike the first-order characterization in Theorem \ref{thm:RobIm}, the equivalence \eqref{eq:ambiguity} is exact: no limit $\Delta\downarrow 0$ is involved. The reason is that $\pi\mapsto \Expect_\pi E(z,x+h)$ is linear in $\pi$, so its infimum over the convex set $\Pi(\Delta)$ is attained at the extreme points, which are precisely the point masses in $\Pi_0(\Delta)$.

\subsection{Generic full column rank of \texorpdfstring{$A_+$}{}}\label{app:full_column_A}

The following proposition shows that the matrix $A_+$ in \eqref{eq:Abplus} generically has full column rank.

\begin{prop}\label{prop:full_column_A}
Let $I\ge J-1$, $\set{h_i}_{i=1}^I$ be the basis functions, and set $h_0\equiv 1$. Suppose that there exist $\set{i_\ell}_{\ell=1}^J\subset \set{0,1,\dots,I}$ with $i_1=0$ and $\set{\tau_j}_{j=1}^J\subset (0,T]$ such that
\begin{enumerate*}
    \item at date $\tau_j$, bond $j$ makes a lump-sum payout $f_j\coloneqq F_j(\tau_j)-F_j(\tau_j-)>0$, and
    \item the $J\times J$ matrix $\tilde{H}=(h_{i_\ell}(\tau_j))$ is invertible.
\end{enumerate*}
Then there exists a closed set $S\subset \R^J$ with Lebesgue measure 0 such that the matrix $A_+$ in \eqref{eq:Abplus} has full column rank whenever $(f_1,\dots,f_J)\notin S$.

If in addition all bonds are zero-coupon bonds, then $A_+$ has full column rank.
\end{prop}

The fact that the set of $(f_1,\dots,f_J)$ for which $A_+$ has rank deficiency is contained in a closed set with Lebesgue measure 0 implies that the set of rank deficiency is nowhere dense (has empty interior). In this sense the rank deficiency of $A_+$ is ``rare''. To prove the result, we need the following lemma.

\begin{lem}\label{lem:invertible_gen}
Let $A,B$ be $N\times N$ matrices and define $\phi:\R^N\to \R$ by $\phi(x)=\det(A\diag(x)+B)$, where $\diag(x)$ denotes the diagonal matrix with diagonal entries $x_1,\dots,x_N$. If $\det A\neq 0$, then for any $c\in \R$ the set
\begin{equation*}
    \phi^{-1}(c)\coloneqq \set{x\in \R^N: \phi(x)=c}
\end{equation*}
is closed and has Lebesgue measure $0$.
\end{lem}

\begin{proof}
Since
\begin{align*}
    \det(A\diag(x)+B)&=\det(A(\diag(x)+A^{-1}B))\\
    &=\det(A)\times \det(\diag(x)+A^{-1}B),
\end{align*}
without loss of generality we may assume that $A$ is the identity matrix. Let $B=(b_{mn})$. That $\phi^{-1}(c)$ is closed is obvious because $\phi$ is continuous.

Let us show by induction on the dimension $N$ that $\phi^{-1}(c)$ is a null set. If $N=1$, then $\phi(x)=x_1+b_{11}$, so $\phi^{-1}(c)=\set{c-b_{11}}$ is a singleton, which is a null set. Suppose the claim holds when $N=n-1$ and consider $n$. Let $B_n$ be the $n\times n$ matrix obtained from the first $n$ rows and columns of $B$, and let
\begin{equation*}
    \phi_n(x_1,\dots,x_n)=\det(\diag(x_1,\dots,x_n)+B_n).
\end{equation*}
Clearly $\phi_n$ is affine in each variable $x_1,\dots,x_n$. Using the Laplace expansion along the $n$-th column, it follows that
\begin{equation*}
    \phi_n(x_1,\dots,x_n)=(x_n+b_{nn})\phi_{n-1}(x_1,\dots,x_{n-1})+\psi_{n-1}(x_1,\dots,x_{n-1})
\end{equation*}
for some function $\psi_{n-1}$ that is affine in each variable $x_1,\dots,x_{n-1}$.

Define the sets $\phi_{n-1}^{-1}(0)\subset \R^{n-1}$ and $G\subset \R^n$ by
\begin{align*}
    \phi_{n-1}^{-1}(0)&\coloneqq \set{(x_1,\dots,x_{n-1}):\phi_{n-1}(x_1,\dots,x_{n-1})=0},\\
    G&\coloneqq \set{(x_1,\dots,x_n):(x_1,\dots,x_{n-1})\notin \phi_{n-1}^{-1}(0), x_n=(c-\psi_{n-1})/\phi_{n-1}-b_{nn}}.
\end{align*}
Then $\phi_n^{-1}(c)\subset (\phi_{n-1}^{-1}(0)\times \R)\cup G$. By the induction hypothesis, $\phi_{n-1}^{-1}(0)$ has measure 0 in $\R^{n-1}$. Since $G$ is the graph of a Borel measurable function, by Fubini's theorem it has measure 0. Therefore $\phi_n^{-1}(c)$ is a null set.
\end{proof}

\begin{proof}[Proof of Proposition \ref{prop:full_column_A}]
Define $\mathbf{h}:[0,T]\to \R^{I+1}$ by $\mathbf{h}(t)=(h_0(t),h_1(t),\dots,h_I(t))'$. Let the $j$-th column vector of $A_+$ be $\mathbf{a}_j=(a_{0j},\dots,a_{Ij})'$. By assumption, bond $j$ pays $f_j>0$ at $\tau_j\in (0,T]$, so it follows from \eqref{eq:aij} that
\begin{equation}
    \mathbf{a}_j=\frac{1}{P(x)}\int_{[0,T] \backslash \set{\tau_j}}\e^{-x(t)}\mathbf{h}(t)\diff F_j(t)+\frac{1}{P(x)}\e^{-x(\tau_j)}f_j\mathbf{h}(\tau_j)\eqqcolon \mathbf{p}_jf_j+\mathbf{q}_j. \label{eq:aj}
\end{equation}
Collecting \eqref{eq:aj} into a matrix, we can write $A_+=P\diag(f)+Q$, where $P,Q$ are matrices with $j$-th column vectors $\mathbf{p}_j,\mathbf{q}_j$ and $f=(f_1,\dots,f_J)$. To show that $A_+$ generically has full column rank, let $\tilde{A}_+$ be the $J\times J$ matrix obtained by taking its $i_\ell$-th row for $\ell=1,\dots,J$. Define $\tilde{P},\tilde{Q}$ similarly. Then $\tilde{A}_+=\tilde{P}\diag(f)+\tilde{Q}$. Since $\mathbf{p}_j=\e^{-x(\tau_j)}\mathbf{h}(\tau_j)/P(x)$, we obtain
\begin{equation*}
    \det \tilde{P}=P(x)^{-J}\left(\prod_{j=1}^J \e^{-x(\tau_j)}\right)\det \tilde{H}\neq 0.
\end{equation*}
Therefore by Lemma \ref{lem:invertible_gen}, $\tilde{A}_+$ is generically invertible, so $A_+$ has generically full column rank.

If in addition all bonds are zero-coupon bonds, then \eqref{eq:aj} reduces to $\mathbf{a}_j=\e^{-x(\tau_j)}f_j\mathbf{h}(\tau_j)/P(x)$, where $\tau_j$ is the maturity. Then $A_+=P\diag(f)$, which has full column rank because $\det \tilde{P}\neq 0$ and $f_j>0$ for all $j$.
\end{proof}

The following example shows that the zero-coupon bond assumption in Proposition \ref{prop:full_column_A} is essential.

\begin{exmp}\label{exmp:singularA}
Suppose $I=J-1=1$ and the basis function is $h_1(t)=t$. Bond 1 is a zero-coupon bond with face value $f_1>0$ and maturity $t_1$. Bond 2 pays $f_2>0$ at $t_2$ and $f_3>0$ at $t_3$. To simplify notation, write $x(t_1)=x_1$ etc. The determinant of the matrix $A_+$ is
\begin{align*}
    \det A_+ &= P(x)^{-2}\det\begin{bmatrix}
    f_1\e^{-x_1} & f_2\e^{-x_2}+f_3\e^{-x_3}\\
    f_1\e^{-x_1}t_1 & f_2\e^{-x_2}t_2+f_3\e^{-x_3}t_3
    \end{bmatrix}\\
    &=P(x)^{-2}f_1\e^{-x_1}\left(f_2\e^{-x_2}(t_2-t_1)+f_3\e^{-x_3}(t_3-t_1)\right).
\end{align*}
Therefore for any $t_2<t_1<t_3$ and $f_3>0$, we have $\det A_+=0$ if and only if
\begin{equation}
    (f_1,f_2)\in \set*{(f_1,f_2)\in\R_{++}^2:f_2=f_3\e^{x_2-x_3}\frac{t_3-t_1}{t_1-t_2}}.\label{eq:rank_def_set}
\end{equation}
The closure of the rank deficiency set \eqref{eq:rank_def_set} is a ray in $\R^2$ and has measure 0.
\end{exmp}

\subsection{Key rate duration matching}\label{sec:KRD}

\begin{figure}[!htb]
\centering
\includegraphics[width=0.9\linewidth]{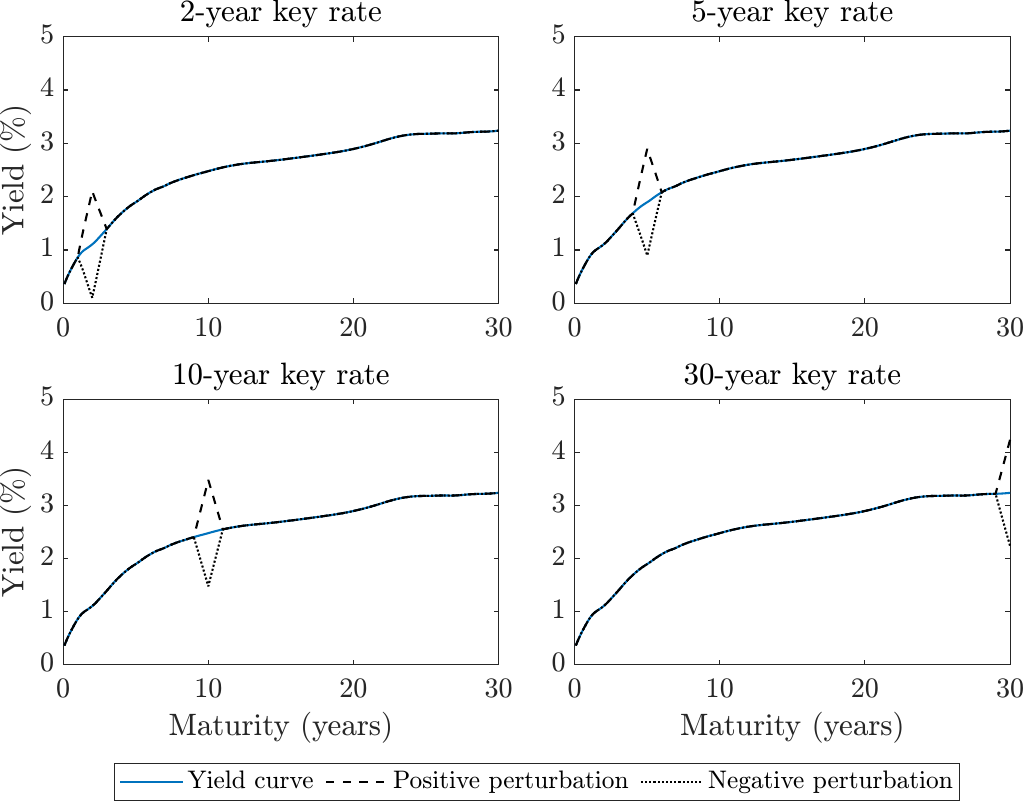}
\caption{Key rate perturbations.}\label{fig:key_perturb}
\caption*{\footnotesize Note: The panels show positive and negative perturbations to the yield curve due to a 1\% change in the respective key rate. We linearly interpolate the yields after a change in the key rate to ensure that the yield curve remains continuous. The true yield curve (in blue) is calculated on December 2, 2016.}
\end{figure}

This appendix explains the key rate duration matching method of \citet{Ho1992}. The key rate duration of a bond with yield curve $y$ and yield change $\Delta$ at time to maturity $t$ is defined by
\begin{equation*}
\mathrm{KRD}(y,t,\Delta)\coloneqq \frac{P(y_-) - P(y_+)}{2 \Delta P(y)},
\end{equation*}
where $y_\pm$ denotes the yield curve after changing $y(t)$ to $y(t)\pm \Delta$ at a specific term $t$ and linearly interpolating between the adjacent terms. Following the literature, we set the shift to $\Delta=0.01$ (100 basis points). Figure \ref{fig:key_perturb} illustrates the procedure for a set of key rates on December 2, 2016.\footnote{The key rate duration of a zero-coupon bond with maturity $t$ is equal to $t$ and zero otherwise. Since we use linear interpolation after a key rate perturbation to keep the yield curve continuous, the key rate for a zero-coupon bond with maturity $t$ is not exactly equal to $t$ in our application.}

In our simulation, we consider five zero-coupon bonds and aim to match the liability’s key rate exposures at the maturities of these bonds. In addition, we impose a value-matching constraint to ensure consistency with other immunization methods. The \KRD portfolio should also satisfy the market frictions we impose in the static and dynamic hedging experiments. This leads to more restrictions than unknowns. To determine the optimal portfolio, we therefore minimize the mean squared distance between the key rate exposures of the bond portfolio and those of the liability, subject to the constraints.

\section{Alternative specifications and data}

\subsection{Alternative basis functions}\label{app:kernel_basis}

As a robustness check on how well the Chebyshev basis captures yield curve changes, we repeat the goodness-of-fit exercise with the kernel basis functions of \citet{filipovic2025stripping}. They measure the smoothness of the curve $g(t) = e^{-x(t)}$, with $x(t)$ denoting the cumulative discount rate, by a weighted average of its first two derivatives,
\begin{equation}
\norm{g}_{\alpha,\delta} = \left( \int_0^\infty \big[\, \delta\, g'(t)^2 + (1-\delta)\, g''(t)^2 \,\big]\, \e^{\alpha t}\, \diff t \right)^{1/2},
\label{eq:fpy_norm}
\end{equation}
where the tension $\delta \in [0,1]$ balances the slope penalty (against oscillations) and the curvature penalty (against kinks), and the maturity weight $\alpha \ge 0$ enters through $\e^{\alpha t}$. \citet{filipovic2025stripping} derive that the estimated $\hat{g}$ can be expressed as a kernel ridge (KR) solution:
\begin{equation*}
\hat{g}(t) = 1 + \sum_{n=1}^N \beta_n k(t,t_n),
\end{equation*}
where for $\delta = 0$
\begin{equation}
k(s,t) = -\frac{s\wedge t}{\alpha^2}\,\e^{-\alpha (s\wedge t)} + \frac{2}{\alpha^3}\Big(1 - \e^{-\alpha (s\wedge t)}\Big) - \frac{s\wedge t}{\alpha^2}\,\e^{-\alpha (s\vee t)},
\label{eq:fpy_kernel}
\end{equation}
with $s\wedge t = \min(s,t)$ and $s\vee t = \max(s,t)$.

\begin{figure}[!htb]
\centering
\includegraphics[width=0.48\linewidth]{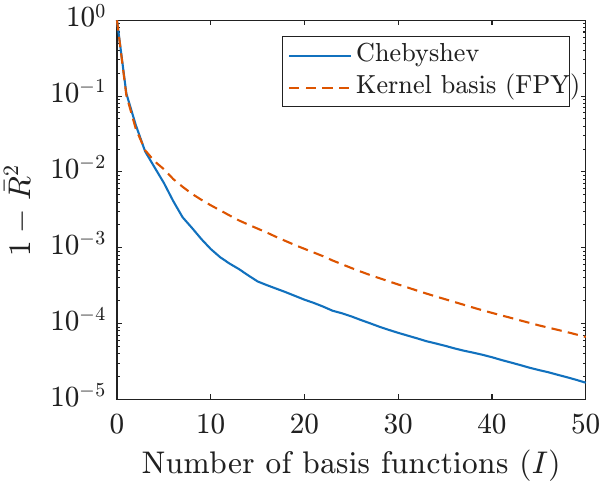}
\caption{Goodness-of-fit under the Chebyshev and \citet{filipovic2025stripping} kernel bases.}\label{fig:R2I_basis}
\caption*{\footnotesize Note: The figure shows the unexplained component $1-\bar{R}^2$ of yield curve changes on the training window (November 1985--December 1995) as the number of basis functions $I$ increases, for the Chebyshev basis and the orthogonalized kernel factors of \citet{filipovic2025stripping}.}
\end{figure}

The functions $k(\cdot,t_n)$ span deviations of the (exponential) cumulative discount curve $g$. As our basis we take the leading $I$ eigenvectors of the Gram matrix $K_{ij}=k(t_i,t_j)$---the orthogonalized KR factors, ordered by decreasing eigenvalue---and compare how well they capture changes in the yield curve relative to our Chebyshev polynomials. Following \citet{filipovic2025stripping} we set $\alpha = 0.05$ and $\delta = 0$. %; the smoothness parameter $\lambda$ rescales only their ridge fit and does not affect the eigenvectors of $K_{ij} = k(t_i,t_j)$. 
Unlike our Chebyshev polynomials, which are ordered by degree, the KR factors are ordered by eigenvalue. By Proposition~\ref{prop:invariance} the robust portfolio depends on the basis only through its span, and Figure~\ref{fig:R2I_basis} shows that the Chebyshev basis attains at least as low an unexplained variance $1-\bar{R}^2$ for every $I$.

\subsection{Static hedging under the \texorpdfstring{$\ell^\infty$}{} norm}\label{app:eval_linf}

The main text reports robust immunization portfolios based on the $\ell^2$ norm, which admit a closed form and are computationally cheap. This appendix repeats the static hedging experiment of Section~\ref{subsec:eval_constr} with the $\ell^\infty$ stress set of Theorem~\ref{thm:RobIm}. The benchmarks (\LSHD, \KRD) are defined as least-squares problems and are unaffected by the norm; their columns are repeated for reference.

Table~\ref{t:constr_linf} reports the results. The $\ell^\infty$ portfolios perform similarly to their $\ell^2$ counterparts, and in most cells slightly better, at markedly lower gross leverage: for example, the unconstrained \RI(2) portfolio for \texttt{fullHorizon} has gross leverage $1.4$ under $\ell^\infty$ against $2.6$ under $\ell^2$, so the leverage and collateral constraints rarely bind. All rankings against the benchmarks are unchanged or improve slightly (\RI(2) edges \LSHD\ for the \texttt{medium} liability under $\ell^\infty$). 

\begin{table}[!htb]
\centering
\caption{Underfunding ratio under portfolio constraints, $\ell^\infty$ norm.}\label{t:constr_linf}
\begin{tabular}{lccccc}
\toprule
Liability type & \RI(0) & \RI(1) & \RI(2) & \LSHD & \KRD \\
\midrule
\multicolumn{6}{c}{\emph{Panel A: no short sales}} \\
\midrule
\texttt{fullHorizon}  & 2.26 & 0.47 & 0.43 & 0.71 & 2.27 \\
\texttt{longRun}      & 7.29 & ---  & ---  & 2.12 & 2.12 \\
\texttt{medium}       & 3.44 & 0.46 & ---  & 0.74 & 5.29 \\
\texttt{shortAndLong} & 1.91 & 0.63 & 0.75$^{\dagger}$ & 1.10 & 0.96 \\
\texttt{pension}      & 1.22 & 0.27 & 0.13 & 0.29 & 2.27 \\
\midrule
\multicolumn{6}{c}{\emph{Panel B: gross-leverage cap}} \\
\midrule
\texttt{fullHorizon}  & 2.28 & 0.86 & 0.44 & 0.71 & 2.27 \\
\texttt{longRun}      & 7.60 & 2.14 & ---  & 2.48 & 1.74 \\
\texttt{medium}       & 3.63 & 0.54 & 0.49 & 0.50 & 5.29 \\
\texttt{shortAndLong} & 1.96 & 0.67 & 0.83 & 1.06 & 0.96 \\
\texttt{pension}      & 1.23 & 0.27 & 0.13 & 0.34 & 2.27 \\
\midrule
\multicolumn{6}{c}{\emph{Panel C: collateral budget}} \\
\midrule
\texttt{fullHorizon}  & 2.28 & 0.86 & 0.44 & 0.66 & 2.27 \\
\texttt{longRun}      & 7.63 & ---  & ---  & 2.12 & 2.12 \\
\texttt{medium}       & 3.63 & 0.54 & 0.49 & 0.53 & 5.29 \\
\texttt{shortAndLong} & 1.96 & 0.67 & 0.91 & 1.09 & 0.96 \\
\texttt{pension}      & 1.23 & 0.27 & 0.13 & 0.43 & 2.27 \\
\bottomrule
\end{tabular}
\caption*{\footnotesize Note: As Table~\ref{t:constr}, with the \RI\ portfolios computed under the $\ell^\infty$ norm. \LSHD\ and \KRD\ are least-squares methods, independent of the norm; their columns are repeated for reference. $^{\dagger}$Feasible on $53\%$ of dates.}
\end{table}

\subsection{Static hedging with the FPY yield curve data}\label{app:eval_FPY}
Here we repeat the static hedging experiment of Section~\ref{subsec:eval_constr} using the yield curve data of \citet{filipovic2025stripping} (hereafter, FPY) instead of \citet{liu2021reconstructing}. We use their daily zero-coupon yields at monthly maturities, trimmed to the same 1996--2023 evaluation sample. Over this sample, the FPY curves are backed by traded securities up to maturities of $23$ to $30$ years on every date (mean $28.4$ years), so the entire bond menu is always observed and only the far end of the 30-year liability horizon relies on their kernel-based extension. All other aspects of the experiment are unchanged.

Table~\ref{t:constr_FPY} reports the results. Every ranking in Table~\ref{t:constr} is preserved: robust immunization attains the best performance for the same liabilities and constraint sets. Underfunding levels are slightly lower throughout for every method, consistent with the smoother FPY curves containing less estimation noise, but no comparison between methods is affected. The conclusions of Section~\ref{subsec:eval_constr} are therefore not driven by features of the \citet{liu2021reconstructing} data.

\begin{table}[!htb]
\centering
\caption{Underfunding ratio under portfolio constraints, FPY data.}\label{t:constr_FPY}
\begin{tabular}{lccccc}
\toprule
Liability type & \RI(0) & \RI(1) & \RI(2) & \LSHD & \KRD \\
\midrule
\multicolumn{6}{c}{\emph{Panel A: no short sales}} \\
\midrule
\texttt{fullHorizon}  & 2.43 & 0.48 & 0.43 & 0.69 & 2.25 \\
\texttt{longRun}      & 6.90 & ---  & ---  & 2.03 & 2.03 \\
\texttt{medium}       & 3.25 & 0.44 & ---  & 0.74 & 5.30 \\
\texttt{shortAndLong} & 2.09 & 0.73 & 0.75$^{\dagger}$ & 1.08 & 0.95 \\
\texttt{pension}      & 1.30 & 0.34 & 0.11 & 0.29 & 2.28 \\
\midrule
\multicolumn{6}{c}{\emph{Panel B: gross-leverage cap}} \\
\midrule
\texttt{fullHorizon}  & 2.45 & 0.92 & 0.49 & 0.71 & 2.25 \\
\texttt{longRun}      & 6.93 & 2.00 & ---  & 2.46 & 1.58 \\
\texttt{medium}       & 3.29 & 0.90 & 0.59 & 0.48 & 5.30 \\
\texttt{shortAndLong} & 2.09 & 1.02 & 0.81 & 1.06 & 0.95 \\
\texttt{pension}      & 1.30 & 0.44 & 0.12 & 0.33 & 2.28 \\
\midrule
\multicolumn{6}{c}{\emph{Panel C: collateral budget}} \\
\midrule
\texttt{fullHorizon}  & 2.45 & 0.99 & 0.53 & 0.66 & 2.25 \\
\texttt{longRun}      & 6.93 & ---  & ---  & 2.03 & 2.03 \\
\texttt{medium}       & 3.29 & 0.99 & 0.59 & 0.50 & 5.30 \\
\texttt{shortAndLong} & 2.09 & 1.05 & 0.91 & 1.09 & 0.95 \\
\texttt{pension}      & 1.30 & 0.44 & 0.12 & 0.42 & 2.28 \\
\bottomrule
\end{tabular}
\caption*{\footnotesize Note: As Table~\ref{t:constr}, using the \citet{filipovic2025stripping} discount bond database. $^{\dagger}$Feasible on $54\%$ of dates.}
\end{table}

\subsection{Dynamic hedging with realized yield curves}\label{app:dyn_hedge_realized}

This appendix repeats the dynamic hedging experiment of Section~\ref{sec:dyn} on realized rather than simulated yield curves. We take every 10-year window of quarterly \citet{liu2021reconstructing} yield curve observations starting between 1996 and 2013, a total of 72 overlapping windows, so the statistics below describe the range of historical experiences rather than a sampling distribution. All other aspects of the experiment (liabilities, constraints, transaction costs, rebalancing) are as in Section~\ref{sec:dyn}.

\begin{figure}[!htb]
\centering
\begin{subfigure}{0.48\linewidth}
    \includegraphics[width=\linewidth]{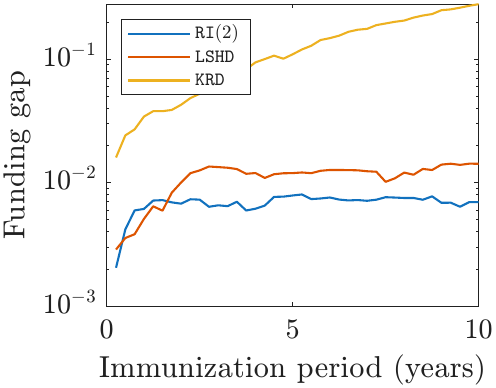}
\caption{\texttt{fullHorizon}.}
\end{subfigure}
\begin{subfigure}{0.48\linewidth}
    \includegraphics[width=\linewidth]{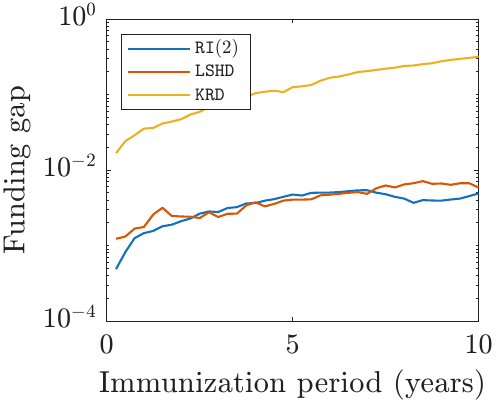}
\caption{\texttt{pension}.}
\end{subfigure}
\caption{Median funding gap on historical yield curve paths.}\label{fig:dynFR}
\caption*{\footnotesize Note: Median absolute funding gap $\abs{1-V_s/P_s}$
across the 72 historical 10-year windows at each point of the immunization
period, on a logarithmic scale. \KRD's gap consists entirely of underfunding:
its funding ratio declines almost linearly to $0.72$ after ten years for
\texttt{fullHorizon}, reflecting its persistent duration gap compounding over
the secular fall in yields. \RI(2) remains within a fraction of a percent of
full funding throughout. \HD\ is omitted; see Table~\ref{t:dyn_realized}.}
\end{figure}

Table~\ref{t:dyn_realized} reports the results, and Figure~\ref{fig:dynFR} shows the median absolute funding gap over the windows. Robust immunization keeps the fund essentially fully funded throughout: for the \texttt{pension} liability, the mean absolute return error after ten years is $0.6\%$ and the worst window since 1996 is $1.8\%$, at a gross leverage of one and transaction costs of a tenth of a basis point per year. \LSHD\ is the closest competitor but uses the full leverage budget and roughly $2.5$ times the turnover to achieve errors up to three times larger; its small signed moment mismatches under the binding cap, which roughly cancel on trendless simulated paths, compound over the secular decline in yields. The same mechanism is an order of magnitude stronger for \KRD: its five-year duration gap (Section~\ref{sec:dyn}), multiplied by the realized decline in yields and never corrected (turnover of $0.1$ per year), accounts for the bulk of its terminal error of $26$--$30\%$, and its median funding ratio declines almost linearly to $0.72$ after ten years for \texttt{fullHorizon}. Consistent with this mechanism, \KRD's worst window begins in 2010, the decade with the largest subsequent decline in yields.

\begin{table}[!htb]
\centering
\caption{Dynamic hedging performance on realized yield curves.}\label{t:dyn_realized}
\begin{tabular}{lcccccc}
\toprule
Method & Mean & p99 & MSE $\times 10^4$ & Leverage & Turnover & Cost (bp/yr) \\
\midrule
\multicolumn{7}{c}{\emph{Panel A: \texttt{fullHorizon}}} \\
\midrule
\RI(2) & 0.99 &  2.82 &   1.7 & 1.30 & 0.40 & 0.20 \\
\LSHD  & 1.85 &  6.07 &   5.4 & 2.00 & 0.75 & 0.43 \\
\KRD   & 25.6 &  40.0 & 719.2 & 1.00 & 0.11 & 0.07 \\
\HD    & 6.00 &  22.8 &  64.6 & 5.99 & 6.24 & 2.56 \\
\midrule
\multicolumn{7}{c}{\emph{Panel B: \texttt{pension}}} \\
\midrule
\RI(2) & 0.64 &  1.78 &   0.7 & 1.00 & 0.19 & 0.10 \\
\LSHD  & 0.72 &  1.91 &   0.7 & 1.99 & 0.50 & 0.26 \\
\KRD   & 29.9 &  44.9 & 966.2 & 1.00 & 0.10 & 0.05 \\
\HD    & 1.61 &  6.88 &   4.6 & 2.14 & 1.30 & 0.59 \\
\bottomrule
\end{tabular}
\caption*{\footnotesize Note: As Table~\ref{t:dyn}, for the 72 overlapping historical 10-year windows starting 1996--2013.}
\end{table}

\end{document}